\documentstyle[aps,preprint,prl,epsf]{revtex}
\tightenlines

\newcommand{\be}{\begin{equation}}
\newcommand{\ee}{\end{equation}}

\begin{document}

\title{\Large{\bf Correlation Effects in Itinerant Magnetism}}

\vskip0.5cm 

\author{ G. G\'{o}rski and J. Mizia }

\address{Institute of Physics, University of Rzesz\'{o}w, ulica Rejtana 16A, \\
35-958 Rzesz\'{o}w, Poland\\}

\maketitle
\vskip0.5cm 
\begin{abstract}

In this contribution we would like to revisit the problems 
of ferromagnetism (F) and antiferromagnetism (AF) 
in the pure itinerant model. These methods can be extended 
later to the superconducting materials. In our model we assume 
the extended Hubbard Hamiltonian. Transition from the paramagnetic 
state to the ordered state of magnetic nature is decided by 
the competition between kinetic and potential energy in which 
there is an increase in the kinetic energy moderated by the inter-site 
interactions, and a decrease in the potential energy. The competition 
between these two energies results in the existence of critical values 
of interactions for creating magnetic alignment. Only when existing 
in a given material interaction exceeds the critical value for 
a given type of ordering we can have the alignment of this type.
 
The influence of inter-site correlation on F and AF in the presence 
of Coulomb on-site correlation is investigated. The well known 
Landau free energy expansion is corrected for both F and AF ordering 
to reflect the presence of critical interaction in the second order term, 
which leads to the magnetic alignment. 

The numerical results show that the inter-site interactions favor F at 
the end of the band and AF at the half-filled point. In both 
cases of F and AF ordering, these interactions lower substantially the 
Curie and Neel's temperature towards experimental data. This helps to remove
the paradox in magnetism that has persisted for a long time.

\end{abstract}

\vskip2.5cm 

{\bf CONTENT}

1. Introduction

2. The Model

\hspace{0.5cm}2.1 Inter-site averages $I_\sigma = \left\langle {c_{i\sigma }^+ c_{j\sigma}}\right\rangle $ 

\hspace{0.5cm}2.2 Band-width, Molecular Field and Electron Occupation

\hspace{1cm}2.2.1 Ferromagnetic State

\hspace{1cm}2.2.2 Antiferromagnetic State, Diagonaliztion 

3. Free Energy, Static Magnetic Susceptibility

4. Ferromagnetism

\hspace{0.5cm}4.1 Onset of Ferromagnetism

\hspace{0.5cm}4.2 Numerical Results for Magnetization and Curie Temperature

5. Antiferromagnetism

\hspace{0.5cm}5.1 Onset of Antiferromagnetism

\hspace{0.5cm}5.2 Numerical Results for Magnetization and Neel's Temperature

6. Conclusion

\vskip2cm 

\noindent{\Large{\bf 1. INTRODUCTION}}

\vskip0.5cm 

In the solids there are many phenomena such as band magnetism, 
metal-insulator transition, high temperature superconductivity, which are 
closely related to the Coulomb correlation. To describe this correlation 
scientists have developed what is known today as the single band Hubbard 
model \cite{1,2,3}. In the original version of this model only the on-site Coulomb 
repulsion, $U = \left( {i,i\left| {1 \mathord{\left/ {\vphantom {1 r}} 
\right. \kern-\nulldelimiterspace} r} \right|i,i} \right)$ was included. The 
use of the simple mean-field approximation for this model led back to the 
Stoner model \cite{4,5,6}, in which the majority and minority spin bands are 
shifted with respect to each other for the quantity $\Delta E = m \cdot U$, 
where $m$ is the magnetization in Bohr's magnetons, and the on-site Coulomb 
constant $U$ plays the role of the exchange interaction. The Coulomb 
constant $U$ resulting from the Stoner condition for creating ferromagnetism is 
large, i.e. of the order of the bandwidth. On one hand it can be justified 
by the existing strong Coulomb interaction, but on the other hand for such a 
strong interaction, one can not use the mean-field approximation.

These limitations of the mean-field approximation started the search for a 
more realistic ferromagnetic solution using higher order approximations, 
e.g. the Hubbard I approximation \cite{1}, coherent potential approximation (CPA) 
\cite{7,8}. While the Hubbard I approximation describes correctly the atomic 
limit ($t = 0)$ it is not correct for the intermediate $U / t$. For these 
values the coherent potential approximation (CPA), which is equivalent to 
Hubbard III decoupling of the Green functions, is better. Unfortunately, 
after a thorough analysis it was concluded that the Coulomb repulsion $U$, 
no matter how strong, does not lead to the ferromagnetic order (see e.g., 
\cite{9,10}). The reason is that it does not yield the spin dependent band shift 
or spin dependent band narrowing necessary for a ferromagnetic ordering.

Both Hubbard I and CPA approximation have two types of particles; electrons 
moving between empty sites and electrons moving between sites occupied by 
electrons of opposite spin, separated from each other by the energy $U$ and 
forming lower and upper Hubbard's sub-bands. This is the two-pole 
approximation in the language of Green functions. Base on this scheme is the 
newer model called ``spectral density approach'' (SDA) \cite{11,12}. The magnetic 
phase diagrams calculated by this method are more realistic and the values 
of the Curie temperatures also make sense. The main advantage of SDA is to 
obtain a spin-dependent band shift necessary to obtain the ferromagnetic 
ordering. On the other side the SDA method is the linear combination of two 
$\delta $ functions leading to the real self-energy, and lacking the 
quasi-particle damping. Therefore, Nolting and co-workers have proposed a 
combination of SDA and CPA called ``modified alloy analogy'' (MAA) \cite{13}. 
This approximation has brought the self-consistent ferromagnetic solutions, 
and the complex values of self-energy. The defect of this method is the 
small range of concentrations for which one has the ferromagnetism and also 
a slightly unclear derivation of the method.

Another approach to solving the Hubbard model is computer simulation. The 
initial attempts of the Monte Carlo simulation applied to the basic Hubbard 
model also did not give the ferromagnetic ground state \cite{14}. More recently, 
the new dynamical mean-field theory (DMFT) \cite{15,16} has been developed for a 
direct computational simulation of systems with correlated electrons on a 
crystal lattice. This method has the exact solution in the non-trivial limit 
of an infinite coordination number \cite{17}. The results have been obtained by 
using quantum Monte-Carlo (QMC) simulation \cite{18,19,20} and the mean-field Green 
function theory. Use of this dynamical mean field theory (DMFT method), has 
introduced a significant progress in the theory of ferromagnetism. The 
results (see e.g. \cite{21,22}) show the existence of ferromagnetism but at much 
lower temperatures than those coming from the Hartree-Fock approximation. 
Such results could remove the problem known as a ``magnetic paradox'', i.e. 
the Curie temperature coming from the interaction constant (e.g. $U)$ - 
fitted to obtain the correct magnetic moment at $T = 0\mbox{ K}$ - as being 
much too high. The later attempts along the computational lines (see 
\cite{16,20,22,23,24,25}) are still inconclusive, mainly for the reason of the limited 
size of the computational systems which seems to be too small too represent 
the bulk materials. 

In recent years there has been a very fruitful analytical development in 
describing the magnetism by extending the Hubbard model and including into 
it the band degeneration and the inter-site Coulomb interactions. The 
influence of inter-site interactions on the band magnetism has been studied 
by many authors \cite{26,27,28,29,30,31,32,33,34,35,36,37}. They have shown how different inter-site 
interactions would affect the ferromagnetism in the presence of the on-site 
Coulomb repulsion. This also will be the subject of the present chapter. 

On the experimental front in the last few years there has been discovery of 
a triplet superconductivity in UGe$_{2}$ \cite{38}, URhGe \cite{39} and ZrZn$_{2}$ 
\cite{40} materials. Although the transition to superconductivity takes place in 
very low temperatures ($\sim $0.8 K), the mechanism of this phenomena is 
very interesting, since the superconducting pairing is (probably) stimulated 
by the Hund's interaction (see e.g. \cite{41}), the same one which stimulates 
weak itinerant ferromagnetism. It was also found that the growing magnetic 
moment would suppress the superconductivity. The explanation of these facts 
would be quite simple; the Hund's interaction stimulates both F and SC 
ordering, but if the ordering of a given type already takes place then it 
precludes the other type of ordering. 

Another type of new phenomenon which includes the band magnetism is the well 
documented coexistence between superconducting ordering and itinerant 
antiferromagnetism in high-$T_{C}$ cuprates, e.g. YBa$_{2}$Cu$_{3}$O$_{6 + 
x}$, La$_{2 - x}$(Sr,Ba)$_{x}$CuO$_{4}$. Since in these compounds the 
magnetic moment is stronger it suppresses the superconductivity until the 
moment itself disappears (see \cite{42,43,44}).

Because of all these reasons of all these theoretical problems and 
experimental facts we would like in this chapter to revisit the problems of 
ferromagnetism and antiferromagnetism in the pure itinerant model. These 
methods can be extended later to the superconducting materials, where the 
magnetic ordering competes rather than cooperates with superconductivity.

\vskip1cm 

\noindent{\Large{\bf 2. THE MODEL}}

\vskip0.5cm 

The general model of itinerant magnetic ordering is based on the extended 
Hubbard quasi-single band Hamiltonian of the following form \cite{33}

\begin{equation}
\label{eq1}
\begin{array}{c}
 H = - \sum\limits_{ < ij > \sigma } {t_{ij}^\sigma \left( {c_{i\sigma }^ + 
c_{j\sigma } + h.c.} \right)} - \mu _0 \sum\limits_i {\hat {n}_i } - 
F\sum\limits_{i\sigma } {n_{i\sigma } \hat {n}_{i\sigma } } + U\sum\limits_i 
{\hat {n}_{i \uparrow } \hat {n}_{i \downarrow } } + V\sum\limits_{ < ij > } 
{\hat {n}_i \hat {n}_j } \\ 
 + J\sum\limits_{ < ij > \sigma ,\sigma '} {c_{i\sigma }^ + c_{j\sigma '}^ + 
c_{i\sigma '} c_{j\sigma } } + J'\sum\limits_{ < ij > } {\left( {c_{i 
\uparrow }^ + c_{i \downarrow }^ + c_{j \downarrow } c_{j \uparrow } + h.c.} 
\right)} \\ 
 \end{array} \quad ,
\end{equation}
where $\mu _0 $ is the chemical potential, $c_{i\sigma }^ + \left( 
{c_{i\sigma } } \right)$ creates (annihilates) the electron with spin 
$\sigma $ on the $i$-th lattice site, $\hat {n}_{i\sigma } = c_{i\sigma }^ + 
c_{i\sigma } $ is the particle number operator for electrons with spin 
$\sigma $ on the $i$-th lattice site, $\hat {n}_i = \hat {n}_{i\sigma } + \hat 
{n}_{i - \sigma } $ is the charge operator, $n_{i\sigma } $ is the average 
number of electrons on sites $i$ with spin $\sigma $. 

In this Hamiltonian we included not only the dominant on-site Coulomb 
correlation, $U$, but also the on-site Hund's field $F$. Such a field can 
exist only as the interaction between different orbitals in a multi-orbital 
model. We assume the single band, but composed from identical orbitals, 
which are fully degenerate i.e. have the same density of states and the same 
electron occupation (see \cite{10}). In such a band the effective exchange field 
can be expressed as; $F = (d - 1) \cdot J_{in} $, where $J_{in} $ is the 
exchange interaction between different orbitals within the same atomic site, 
and $d$ is the number of orbitals within the band. As a result our model is 
a quasi-single band model. The intra-atomic Hund field in Eq. (\ref{eq1}) is already 
expressed in the Hartree-Fock approximation, which will be justified only 
for small values of this interaction. In addition we have three explicit 
inter-site interactions \cite{1,45,46}; $J$-exchange interaction, $J'$-pair 
hopping interaction, $V$-density-density interaction. To avoid a large 
number of free parameters we will assume later on in the numerical analysis 
that $J'=J,V = 0$, which will leave us with only two parameters; $J$ and 
parameter $S$ representing kinetic interactions (see Eq. (\ref{eq5})).

The spin dependent correlation hopping $t_{ij}^\sigma $ depends on the 
occupation of sites $i$ and$ j$, and in the operator form can be expressed as 

\begin{equation}
\label{eq2}
t_{ij}^\sigma = t(1 - \hat {n}_{i - \sigma } )(1 - \hat {n}_{j - \sigma } ) 
+ t_1 \left[ {\hat {n}_{i - \sigma } (1 - \hat {n}_{j - \sigma } ) + \hat 
{n}_{j - \sigma } (1 - \hat {n}_{i - \sigma } )} \right] + t_2 \hat {n}_{i - 
\sigma } \hat {n}_{j - \sigma } \quad ,
\end{equation}
where $t$ is the hopping amplitude for an electron of spin $\sigma $ when 
both sites $i $and $j$ are empty. Parameters; $t_1 $, $t_2 $, are the hopping 
amplitudes for an electron of spin $\sigma $ when one or both of the sites 
$i$ or $j$ are occupied by an electron with opposite spin, respectively. Quite 
recently, several authors have suggested that the expected relation; $t > 
t_1 > t_2 $, may be reversed for large enough inter-atomic distances; $t < 
t_1 < t_2 $ (see Refs. \cite{47} and \cite{48}). This concept would fit the results of 
Gunnarsson and Christensen \cite{49}, who for the heavier elements (e.g. 3d or 
4f) claim growing hopping integrals with increasing occupation.

Including the occupationally dependent hopping given by Eq. (\ref{eq2}) into the 
Hamiltonian (\ref{eq1}) we obtain the following result 

\begin{equation}
\label{eq3}
\begin{array}{c}
 H = - \sum\limits_{ < ij > \sigma } {\left[ {t - \Delta t\left( {\hat 
{n}_{i - \sigma } + \hat {n}_{j - \sigma } } \right) + 2t_{ex} \hat {n}_{i - 
\sigma } \hat {n}_{j - \sigma } } \right]\left( {c_{i\sigma }^ + c_{j\sigma 
} + h.c.} \right)} - \mu _0 \sum\limits_i {\hat {n}_i } - 
F\sum\limits_{i\sigma } {n_{i\sigma } \hat {n}_{i\sigma } } \\ 
 + U\sum\limits_i {\hat {n}_{i \uparrow } \hat {n}_{i \downarrow } } + 
V\sum\limits_{ < ij > } {\hat {n}_i \hat {n}_j } + J\sum\limits_{ < ij > 
\sigma ,\sigma '} {c_{i\sigma }^ + c_{j\sigma '}^ + c_{i\sigma '} c_{j\sigma 
} } + J'\sum\limits_{ < ij > } {\left( {c_{i \uparrow }^ + c_{i \downarrow 
}^ + c_{j \downarrow } c_{j \uparrow } + h.c.} \right)} \\ 
 \end{array} \quad ,
\end{equation}
where

\begin{equation}
\label{eq4}
\Delta t = t - t_1 \quad ,
\quad
t_{ex} = \frac{t + t_2 }{2} - t_1 \quad .
\end{equation}

In this form it is quite visible that the kinetic interactions: the hopping 
interaction, $\Delta t$, and the exchange hopping interaction, $t_{ex} $, 
are also the inter-site interactions.

We assume that ${t_1 } \mathord{\left/ {\vphantom {{t_1 } t}} \right. 
\kern-\nulldelimiterspace} t = S$ and ${t_2 } \mathord{\left/ {\vphantom 
{{t_2 } {t_1 }}} \right. \kern-\nulldelimiterspace} {t_1 } = S_1 $. In 
general these parameters are different and they both fulfill the condition 
$S < 1$ and $S_1 < 1$ which is equivalent to $t > t_1 > t_2 $ (see Ref. 
\cite{47}). For simplicity we will assume now additionally that $S \equiv S_1 $. 
With this assumption we have the relationships 

\begin{equation}
\label{eq5}
\Delta t = t - t_1 = t(1 - S) \quad ,
\quad
t_{ex} = \frac{t + t_2 }{2} - t_1 = \frac{t}{2}(1 - S)^2 \quad .
\end{equation}

In the Hamiltonian (\ref{eq3}) there are many inter-site interactions: $\Delta 
t,t_{ex} ,J,J',V$, for which we will use the modified Hartree-Fock (H-F) 
approximation. For the two- and three-body terms of Hamiltonian we will 
adopt the procedure introduced by Foglio and Falicov \cite{50}, Aligia and 
co-workers \cite{51}, and Hirsch \cite{27}. Neglecting superconductive averages of the 
type: $\left\langle {c_{i\sigma }^+ c_{j\sigma'}^+} \right\rangle $, 
$\left\langle {c_{i\sigma} c_{j\sigma'}} \right\rangle $, and the 
spin-flip terms $\left\langle {c_{i\sigma }^+ c_{j -\sigma }} 
\right\rangle$ we obtain 

- e.g. for two-body term 

\begin{equation}
\label{eq6}
\begin{array}{c}
 \hat {n}_i \hat {n}_j = \left( {\hat {n}_{i \uparrow } + \hat {n}_{i 
\downarrow } } \right)\left( {\hat {n}_{j \uparrow } + \hat {n}_{j 
\downarrow } } \right) \cong \left\langle {\hat {n}_{i \uparrow } + \hat 
{n}_{i \downarrow } } \right\rangle \left( {\hat {n}_{j \uparrow } + \hat 
{n}_{j \downarrow } } \right) + \left( {\hat {n}_{i \uparrow } + \hat {n}_{i 
\downarrow } } \right)\left\langle {\hat {n}_{j \uparrow } + \hat {n}_{j 
\downarrow } } \right\rangle \\ 
 - \sum\limits_\sigma {\left( {\left\langle {c_{i\sigma }^ + c_{j\sigma } } 
\right\rangle c_{j\sigma }^ + c_{i\sigma } + h.c.} \right)} + const \\ 
 \end{array} \quad ,
\end{equation}

- e.g. for three-body term

\begin{equation}
\label{eq7}
\begin{array}{c}
 c_{i \uparrow }^ + c_{j \uparrow } \hat {n}_{i \downarrow } \hat {n}_{j 
\downarrow } = c_{i \uparrow }^ + c_{j \uparrow } c_{i \downarrow }^ + c_{i 
\downarrow } c_{j \downarrow }^ + c_{j \downarrow } \cong c_{i \uparrow }^ + 
c_{j \uparrow } \left( {\left\langle {\hat {n}_{i \downarrow } } 
\right\rangle \left\langle {\hat {n}_{j \downarrow } } \right\rangle - 
\left\langle {c_{i \downarrow }^ + c_{j \downarrow } } \right\rangle 
\left\langle {c_{j \downarrow }^ + c_{i \downarrow } } \right\rangle } 
\right) \\ 
 - c_{i \downarrow }^ + c_{j \downarrow } \left\langle {c_{i \uparrow }^ + 
c_{j \uparrow } } \right\rangle \left\langle {c_{j \downarrow }^ + c_{i 
\downarrow } } \right\rangle - c_{j \downarrow }^ + c_{i \downarrow } 
\left\langle {c_{i \uparrow }^ + c_{j \uparrow } } \right\rangle 
\left\langle {c_{i \downarrow }^ + c_{j \downarrow } } \right\rangle \\ 
 + \hat {n}_{i \downarrow } \left\langle {c_{i \uparrow }^ + c_{j \uparrow } 
} \right\rangle \left\langle {\hat {n}_{j \downarrow } } \right\rangle + 
\hat {n}_{j \downarrow } \left\langle {c_{i \uparrow }^ + c_{j \uparrow } } 
\right\rangle \left\langle {\hat {n}_{i \downarrow } } \right\rangle + const 
\\ 
 \end{array} \quad .
\end{equation}

The main point of this approximation is to retain the following inter-site 
averages: $I_\sigma = \left\langle {c_{i\sigma }^ + c_{j\sigma } } 
\right\rangle $, in addition to the usual on site averages: $n_{i\sigma } = 
\left\langle {\hat {n}_{i\sigma } } \right\rangle $, which contribute to the 
Stoner field.

After performing this modified Hartree-Fock approximation on all inter-site 
interactions: $\Delta t,t_{ex} ,J,J',V$, we obtain the following simplified 
Hamiltonian

\begin{equation}
\label{eq8}
H = - \sum\limits_{ < ij > \sigma } {t_{eff}^\sigma \left( {c_{i\sigma }^ + 
c_{j\sigma } + h.c.} \right)} - \mu _0 \sum\limits_i {\hat {n}_i } + 
\sum\limits_{i\sigma } {M_i^\sigma \hat {n}_{i\sigma } } + U\sum\limits_i 
{\hat {n}_{i \uparrow } \hat {n}_{i \downarrow } } \quad ,
\end{equation}
where $t_{eff}^\sigma = t \cdot b^\sigma $ is the effective hopping 
integral, with $b^\sigma $given by

\begin{equation}
\label{eq9}
b^\sigma = 1 - \frac{\Delta t}{t}\left( {n_{i - \sigma } + n_{j - \sigma } } 
\right) + 2\frac{t_{ex} }{t}\left( {n_{i - \sigma } n_{j - \sigma } - I_{ - 
\sigma }^2 - 2I_\sigma I_{ - \sigma } } \right) - \frac{J - V}{t}I_\sigma - 
\frac{J + J'}{t}I_{ - \sigma } \quad ,
\end{equation}
and $M_i^\sigma $ is the spin-dependent modified molecular field for 
electrons with spin $\sigma $ expressed as 

\begin{equation}
\label{eq10}
M_i^\sigma = - Fn_{i\sigma} - J\sum\limits_j {n_{j\sigma }}  + 
V\sum\limits_j {\left( {n_{j\sigma } + n_{j - \sigma } } \right)} + 
2z\Delta tI_{ - \sigma } - 2t_{ex} I_{ - \sigma } \sum\limits_j 
{n_{j\sigma } } \quad ,
\end{equation}
where $z$ is the number of the nearest neighbors, $\sum\nolimits_j$ is the 
sum over nearest neighbors of the lattice site $i$. The above molecular field 
is the sum of the on-site contribution $F$ and the inter-site contributions, 
which will be different for F and AF. 

\vskip0.5cm 

{\bf 2.1 INTER-SITE AVERAGES }$I_\sigma = \left\langle {c_{i\sigma }^ + 
c_{j\sigma } } \right\rangle $

\vskip0.5cm 

Parameter $I_\sigma $ can be calculated from the average kinetic energy. In 
the case of ferromagnetism one can write the following expression for the 
average kinetic energy of $ + \sigma $ electrons $\left\langle {K^\sigma } 
\right\rangle $ 

\begin{equation}
\label{eq11}
\left\langle {K^\sigma } \right\rangle = - t_{eff}^\sigma \sum\limits_{ < ij 
> } {\left\langle {c_{i\sigma }^ + c_{j\sigma } } \right\rangle } = - 
zt_{eff}^\sigma I_\sigma = - D_{eff}^\sigma I_\sigma \quad ,
\end{equation}
where $D_{eff}^\sigma = zt_{eff}^\sigma $ is the half band-width of the $ + 
\sigma $ electrons.

The kinetic energy can be calculated straightforward as 

\begin{equation}
\label{eq12}
\left\langle {K^\sigma } \right\rangle = \int\limits_{ - D}^D {f\left[ 
{\varepsilon ^\sigma \left( {\varepsilon ^0} \right)} \right] \cdot 
\varepsilon ^\sigma \left( {\varepsilon ^0} \right) \cdot \rho ^0\left( 
{\varepsilon ^0} \right)d\varepsilon ^0 = - D_{eff}^\sigma I_\sigma } \quad 
,
\end{equation}
with $\varepsilon ^\sigma \left( {\varepsilon ^0} \right) = b^\sigma 
\varepsilon ^0$ being the energy of the deformed band, $\varepsilon ^0$ is 
the energy of unperturbed band with all inter-site interactions being equal 
to zero, and $f\left[ {\varepsilon ^\sigma (\varepsilon ^0)} \right] = 
\frac{1}{1 + e^{{\left[ {\varepsilon ^\sigma (\varepsilon ^0) - \mu } 
\right]} \mathord{\left/ {\vphantom {{\left[ {\varepsilon ^\sigma 
(\varepsilon ^0) - \mu } \right]} {kT}}} \right. \kern-\nulldelimiterspace} 
{kT}}}$ is the Fermi function. Eq. (\ref{eq12}) means that we can calculate the 
average product of two nearest neighbor operators: $I_\sigma = - 
\left\langle {K^\sigma } \right\rangle / D_{eff}^\sigma \,$. 

In the case of zero temperature, $T = 0\mbox{ K}$, and constant density of 
state (DOS): $\rho ^0\left( {\varepsilon ^0} \right) = const = \frac{1}{D}$ 
for $ - D \le \varepsilon ^0 \le D$ ($D = zt$ - unperturbed half-bandwidth), 
we obtain from Eq. (\ref{eq12}) that

\begin{equation}
\label{eq13}
I_\sigma = n_\sigma \left( {1 - n_\sigma } \right) \quad .
\end{equation}

This approximation suggests that we can treat the average $I_\sigma = 
\left\langle {c_{i\sigma }^ + c_{j\sigma } } \right\rangle $ as the 
probability of electron with spin $\sigma $ hopping from the $i$ to the $j$ 
lattice site and back. More precisely, it is given by the average of two 
products. One is the product of probabilities that there is an electron with 
spin $\sigma $ on the $i$ site and that the $j$ site has empty states; 
$\frac{\tilde {n}_{i\sigma } (\tilde {n}_{j\sigma }^t - \tilde {n}_{j\sigma 
} )}{\tilde {n}_{i\sigma } + (\tilde {n}_{j\sigma }^t - \tilde {n}_{j\sigma 
} )}$, and the second one is the probability of the opposite jump; 
$\frac{\tilde {n}_{j\sigma } (\tilde {n}_{i\sigma }^t - \tilde {n}_{i\sigma 
} )}{\tilde {n}_{j\sigma } + (\tilde {n}_{i\sigma }^t - \tilde {n}_{i\sigma 
} )}$. The quantity $\tilde {n}_{i\sigma}$ is the average number of 
electrons in the sub-band $\sigma $ on sites $i$ and $\tilde {n}_{i\sigma }^t $ 
is the total capacity of the sub-band $\sigma $ on site $i$. The total hopping 
probability, called $I_\sigma $, is given by the average of above 
probabilities

\begin{equation}
\label{eq14}
I_\sigma \equiv \frac{1}{2}\left[ {\frac{\tilde {n}_{i\sigma } (\tilde 
{n}_{j\sigma }^t - \tilde {n}_{j\sigma } )}{\tilde {n}_{i\sigma } + (\tilde 
{n}_{j\sigma }^t - \tilde {n}_{j\sigma } )} + \frac{\tilde {n}_{j\sigma } 
(\tilde {n}_{i\sigma }^t - \tilde {n}_{i\sigma } )}{\tilde {n}_{j\sigma } + 
(\tilde {n}_{i\sigma }^t - \tilde {n}_{i\sigma } )}} \right] \quad .
\end{equation}

For the weak correlation ($U < < D)$, when the band is not split by the 
Coulomb repulsion, we have $\tilde {n}_{i\sigma }^t = 1$ and $\tilde 
{n}_{i\sigma } = n_{i\sigma } $

\begin{equation}
\label{eq15}
I_\sigma \equiv \frac{1}{2}\left[ {\frac{n_{i\sigma } (1 - n_{j\sigma } 
)}{n_{i\sigma } + (1 - n_{j\sigma } )} + \frac{n_{j\sigma } (1 - n_{i\sigma 
} )}{n_{j\sigma } + (1 - n_{i\sigma } )}} \right] \quad .
\end{equation}

Expressions (\ref{eq14}) and (\ref{eq15}) give a general form of quantity $I_\sigma $, which 
depend on the type of magnetic ordering and on the strength of Coulomb 
on-site interaction through the mean values $\tilde {n}_{i\sigma } $ and 
$\tilde {n}_{j\sigma } $. 

At first let us consider the quantity $I_\sigma $ in the {\bf ferromagnetic} 
state when one has the obvious relation; $\tilde {n}_{i\sigma } = \tilde 
{n}_{j\sigma } $. In the case of a weak correlation ($U < < D)$ the band is 
not split and for the electrons with spin $\sigma $ we have; $\tilde 
{n}_{i\sigma } = n_{i\sigma } \equiv n_\sigma $, and expression (\ref{eq15}) can be 
simplified to Eq. (\ref{eq13}). 

In the case of strong correlation ($U > > D)$ when the band is split into 
the lower and the upper Hubbard sub-band, quantities $\tilde {n}_{i\sigma }$ and $\tilde {n}_{i\sigma}^t$, will depend 
on the location of the chemical potential. When $n < 1$ the chemical potential is located in the 
lower Hubbard sub-band and we have; $\tilde {n}_{i\sigma } = n_{\sigma}$, 
$\tilde {n}_{i\sigma }^t = 1 - n_{-\sigma}$, what gives for $I_\sigma $ 
the result 

\begin{equation}
\label{eq16}
I_\sigma = \frac{n_\sigma \left( {1 - n} \right)}{1 - n_{ - \sigma } } \quad 
,
\end{equation}
-for the upper Hubbard sub-band ($n > 1)$ we have: $\tilde {n}_{i\sigma } = 
n_\sigma - (1 - n_{ - \sigma } ) = n - 1$, $\tilde {n}_{i\sigma }^t = n_{ - 
\sigma } $ and $I_\sigma $ is equal to 

\begin{equation}
\label{eq17}
I_\sigma = \frac{\left( {n - 1} \right)\left( {1 - n_\sigma } \right)}{n_{ - 
\sigma } } \quad .
\end{equation}

In the {\bf antiferromagnetic} state the crystal lattice will be divided 
into two interpenetrating sub-lattices; $\alpha ,\beta $ with opposite 
spins, and with the average electron numbers equal to

\begin{equation}
\label{eq18}
n_{\pm \sigma }^\alpha = n_{\pm \sigma } = \frac{n\pm m}{2} \quad ,
\quad
n_{\pm \sigma }^\beta = n_{ \mp \sigma } = \frac{n \mp m}{2} \quad ,
\end{equation}
where $m$ is the antiferromagnetic moment per atom in Bohr's magnetons.

The indices $i, j$ belong to the neighboring sub-lattices $\alpha ,\beta $ with 
opposite magnetic moments. In the result for AF ground state the parameter; 
$I_\sigma = I_{ - \sigma } \equiv I_{AF} $ is spin independent (i.e. it does 
not depend on the first power of antiferromagnetic moment $m)$

The parameter $I_{AF} $ for the weak correlation is calculated from Eq. (\ref{eq15}) 
using values of $n_\sigma ^{\alpha \left( \beta \right)} $ from Eq. (\ref{eq18}) 

\begin{equation}
\label{eq19}
I_{AF} = \frac{2n - n^2 - m^2}{4(1 - m^2)} \quad .
\end{equation}

For the strong correlation and the chemical potential located in the lower 
sub-band $\left( {n \le 1} \right)$, we obtain from Eq. (\ref{eq14}) that

\begin{equation}
\label{eq20}
I_{AF} = \frac{\left( {2n - n^2 - m^2} \right)\left( {1 - n} \right)}{(2 - 
n)^2 - m^2} \quad ,
\end{equation}
and for the upper sub-band

\begin{equation}
\label{eq21}
I_{AF} = \frac{\left( {n - 1} \right)\left( {2n - n^2 - m^2} \right)}{n^2 - 
m^2} \quad .
\end{equation}

After calculating the quantity $I_\sigma $ for F and $I_{AF} $ for AF 
ordering we can find the bandwidth parameter, $b^\sigma $, and the modified 
molecular field, $M_i^\sigma $, which are present in the Hamiltonian of Eq. 
(\ref{eq8}).

\vskip0.5cm 

{\bf 2.2 BAND-WIDTH, MOLECULAR FIELD AND ELECTRON OCCUPATION}

\vskip0.5cm 

{\bf 2.2.1 FERROMAGNETIC STATE}

\vskip0.5cm 

In the {\bf ferromagnetic} state the electron concentration is the same on 
each lattice site: $n_{i\sigma } = n_{j\sigma } = n_\sigma $, and the spin 
dependent parameter $I_\sigma $ in the case of weak Coulomb correlation is 
given by Eq. (\ref{eq13}), and Eqs (\ref{eq16}) or (\ref{eq17}) in the case of the strong 
correlation. Inserting those values into Eqs (\ref{eq9}) and (\ref{eq10}) we have 

\begin{equation}
\label{eq22}
b^\sigma = 1 - 2\frac{\Delta t}{t}n_{ - \sigma } + 2\frac{t_{ex} }{t}\left( 
{n_{ - \sigma }^2 - I_{ - \sigma }^2 - 2I_\sigma I_{ - \sigma } } \right) - 
\frac{J - V}{t}I_\sigma - \frac{J + J'}{t}I_{ - \sigma } \quad ,
\end{equation}

\begin{equation}
\label{eq23}
M_i^\sigma \equiv M_{\sigma} = - (F + zJ)n_\sigma + zVn + 2zI_{ - \sigma } 
\left( {\Delta t - t_{ex} n_\sigma } \right) \quad .
\end{equation}

Using Eq. (\ref{eq5}) we can simplify Eq. (\ref{eq22}) to the form

\begin{equation}
\label{eq24}
b^\sigma = 1 - 2\left( {1 - S} \right)n_{ - \sigma } + \left( {1 - S} 
\right)^2\left( {n_{ - \sigma }^2 - I_{ - \sigma }^2 - 2I_\sigma I_{ - 
\sigma } } \right) - \frac{J - V}{t}I_\sigma - \frac{J + J'}{t}I_{ - \sigma } \quad .
\end{equation}

The effective total field in a ferromagnetic state, $F_{tot}^F $,can be 
found now from the Weiss assumption that the energy shift between both spin 
sub-bands, $\Delta E$, is equal to this field multiplied by the existing 
magnetic moment

\begin{equation}
\label{eq25}
\Delta E = M^{ - \sigma } - M^\sigma = F_{tot}^F \cdot m \quad ,
\end{equation}
which after assuming the Eq. (\ref{eq13}) for $I_\sigma $ gives 

\begin{equation}
\label{eq26}
F_{tot}^F = F + z[J + t_{ex} \frac{n^2 - m^2}{2} + 2\Delta t(1 - n)] \quad .
\end{equation}

In analyzing the Hamiltonian (\ref{eq8}) for the appearance of ferromagnetism we 
will use the level of approximation which will modify the DOS, in addition 
to its shift by the molecular field. 

After Fourier transform of the kinetic energy the Hamiltonian (\ref{eq8}) takes on 
the following form

\begin{equation}
\label{eq27}
H = \sum\limits_{k\sigma } {(\varepsilon _k^\sigma + M^\sigma - \mu )\hat 
{n}_{k\sigma } } + U\sum\limits_{i\sigma } {\hat {n}_{i\sigma } \hat {n}_{i 
- \sigma } } 
\end{equation}
with the spin dependent electron dispersion relation given by 

\begin{equation}
\label{eq28}
\varepsilon _k^\sigma = \varepsilon _k^0 b^\sigma \quad ,
\end{equation}
where $b^\sigma $ and $M^\sigma $for the ferromagnetic state are given by 
Eqs (\ref{eq22}) and (\ref{eq23}), and $\varepsilon _k^0 $ is the initial dispersion energy 
of the electron (without the inter-site interactions) 

\begin{equation}
\label{eq29}
\varepsilon _k^0 = - t\sum\limits_{ < i,j > } {e^{i{\rm {\bf k}}\left( {{\rm 
{\bf R}}_{\rm {\bf i}} - {\rm {\bf R}}_{\rm {\bf j}} } \right)}} \quad .
\end{equation}

The Hamiltonian (\ref{eq27}) is solved in the CPA approximation (see \cite{8,52}), which 
produces the following equation for the on-site self-energy $\Sigma _\sigma 
\left( \varepsilon \right)$ 

\begin{equation}
\label{eq30}
(1 - n_{ - \sigma } )\frac{ - \Sigma _\sigma }{1 + \Sigma _\sigma F_\sigma 
(\varepsilon )} + n_{ - \sigma } \frac{U - \Sigma _\sigma }{1 - (U - \Sigma 
_\sigma )F_\sigma (\varepsilon )} = 0 \quad ,
\end{equation}
with the spin dependent Slater-Koster function $F_\sigma (\varepsilon )$ 
which can be written as 

\begin{equation}
\label{eq31}
F_\sigma (\varepsilon ) = \frac{1}{N}\sum\limits_k {\frac{1}{\varepsilon - 
\varepsilon _k^\sigma - M^\sigma - \Sigma _\sigma + \mu }} \quad .
\end{equation}

This function $F_\sigma (\varepsilon )$ can be expressed by the unperturbed 
function 

\begin{equation}
\label{eq32}
F_0 (\varepsilon ) = \frac{1}{N}\sum\limits_k {\frac{1}{\varepsilon - 
\varepsilon _k^0 }} \quad ,
\end{equation}
by the help of the following relation

\begin{equation}
\label{eq33}
F_\sigma \left( \varepsilon \right) = \frac{1}{b^\sigma }F_0 \left( 
{\frac{\varepsilon - M^\sigma + \mu - \Sigma _\sigma }{b^\sigma }} \right) 
\quad ,
\end{equation}
which becomes the standard CPA relation: $F_\sigma \left( \varepsilon 
\right) = F_0 \left( {\varepsilon - M^\sigma + \mu - \Sigma _\sigma } 
\right)$, when all the inter-site interactions are zero and as a result 
$b^\sigma \equiv 1$. The above defined Slater-Koster function was obtained 
from the solution of the CPA equation and depends on the spin orientation 
and on the on-site interaction, $U$, and the inter-site interactions. 

For the unperturbed Slater-Koster function we have the relation \cite{8}

\begin{equation}
\label{eq34}
\rho (\varepsilon ) = - \frac{1}{\pi }\rm Im F_0 (\varepsilon ) \quad .
\end{equation}

Using the identity 

\begin{equation}
\label{eq35}
\rho _\sigma (\varepsilon ) = - \frac{1}{\pi }\rm Im F_\sigma (\varepsilon )
\end{equation}
one can calculate the perturbed DOS depending on the on-site Coulomb 
correlation $U$ through the self-energy $\Sigma _\sigma $, and on the 
inter-site interactions through the band-width parameter $b^\sigma $. 

Using this DOS one can write the expression for electron occupation in the 
ferromagnetic state

\begin{equation}
\label{eq36}
n_{\pm \sigma } = \int\limits_{ - \infty }^\infty {\rho _{^{\pm \sigma }} 
\left( \varepsilon \right)\frac{d\varepsilon }{1 + e^{{\left( {\varepsilon - 
\mu + M^{\pm \sigma }} \right)} \mathord{\left/ {\vphantom {{\left( 
{\varepsilon - \mu + M^{\pm \sigma }} \right)} {kT}}} \right. 
\kern-\nulldelimiterspace} {kT}}}} \quad ,
\end{equation}
where $\rho _\sigma \left( \varepsilon \right)$ is the spin dependent 
density of states given by Eq. (\ref{eq35}), and the previous CPA equations.

The electron occupation number and the magnetization are given by 

\begin{equation}
\label{eq37}
n = n_\sigma + n_{ - \sigma } \quad ,
\end{equation}

\begin{equation}
\label{eq38}
m = n_\sigma - n_{ - \sigma } \quad .
\end{equation}

\vskip0.5cm 

{\bf 2.2.2 ANTIFERROMAGNETIC STATE, DIAGONALIZATION}

\vskip0.5cm 

In the {\bf antiferromagnetic} state the magnetic moment on the nearest 
lattice sites is opposite with respect to each other (see Eqs (\ref{eq18})), the 
quantity$I_\sigma = I_{ - \sigma } = I_{AF} $, and the bandwidth reduction 
parameter $b^\sigma $ from Eq. (\ref{eq9}) is spin independent 

\begin{equation}
\label{eq39}
b^\sigma = b^{ - \sigma } = b^{AF} = 1 - \frac{\Delta t}{t}n + 2\frac{t_{ex} 
}{t}\left( {\frac{n^2 - m^2}{4} - 3I_{AF}^2 } \right) - \frac{2J + J' - 
V}{t}I_{AF} \quad .
\end{equation}

After expressing the kinetic terms by a factor $S$ defined in Eq. (\ref{eq5}) it 
takes on the form

\begin{equation}
\label{eq40}
b^{AF} = 1 - \left( {1 - S} \right)n + \left( {1 - S} \right)^2\left( 
{\frac{n^2 - m^2}{4} - 3I_{AF}^2 } \right) - \frac{2J + J' - V}{t}I_{AF} 
\quad .
\end{equation}

The generalized (modified) molecular field from Eq. (\ref{eq10}) depends on the spin 
and on the sub-lattice index: $\alpha $ or $\beta $. For the sub-lattice 
$\alpha $ we have 

\begin{equation}
\label{eq41}
M_\alpha ^\sigma = - Fn_\sigma - z\left( {J + 2t_{ex} I_{AF} } \right)n_{ - 
\sigma } + zVn + 2z\Delta tI_{AF} \quad ,
\end{equation}
and for the sub-lattice $\beta $

\begin{equation}
\label{eq42}
\begin{array}{c}
 M_\beta ^\sigma = - Fn_{ - \sigma } - z\left( {J + 2t_{ex} I_{AF} } 
\right)n_\sigma + zVn + 2z\Delta tI_{AF} \\ 
 M_\beta ^\sigma \equiv M_\alpha ^{ - \sigma } \\ 
 \end{array} \quad .
\end{equation}

Now, we can calculate the effective total field on the sites $\alpha ,\beta 
$ in a similar way to the case of ferromagnetism

\begin{equation}
\label{eq43}
F_{tot}^{AF} = \frac{M_\alpha ^{ - \sigma } - M_\alpha ^\sigma }{m} = 
\frac{M_\beta ^\sigma - M_\beta ^{ - \sigma } }{m} = F - z\left( {J + 
2t_{ex} I_{AF} } \right) \quad .
\end{equation}

This equation shows that the positive inter-site exchange interaction, $J$, and 
exchange-hopping interaction, $t_{ex} $, are opposing AF. It is contrary to 
the case of ferromagnetism (see Eq. (\ref{eq26})), where both these interactions 
(when positive) are helping the ferromagnetism. 

As mentioned above we assume the type of the crystal lattice, which can be 
divided into two interpenetrating sub-lattices: $\alpha ,\beta $, with the 
average electron numbers given by Eqs (\ref{eq18}).

For the AF ground state we use the diagonalization of Plischke {\&} Mattis 
\cite{53}, Brouers \cite{54} and Mizia \cite{55}. The Hamiltonian (\ref{eq8}), after taking into 
account Eqs (\ref{eq39})-(\ref{eq42}), will take on the following form 

\begin{equation}
\label{eq44}
H = - t_{eff} \sum\limits_{i\sigma } {\left( {\alpha _{i\sigma }^ + \beta 
_{i\sigma } + \beta _{i\sigma }^ + \alpha _{i\sigma } } \right)} + 
\sum\limits_{\begin{array}{c}
 i,\gamma, \sigma \\(\gamma = \alpha ,\beta ) \\ 
 \end{array}} {\left(M_\gamma ^\sigma - \mu _0 +U \hat {n}_{i - \sigma }^\gamma \right) \hat {n}_{i\sigma }^\gamma }
\quad ,
\end{equation}
where $\alpha _{i\sigma }^ + (\alpha _{i\sigma } )$ and $\beta _{i\sigma }^ 
+ (\beta _{i\sigma } )$ are the creation (annihilation) operators for an 
electron of spin $\sigma $ on the sub-lattice $\alpha $ and $\beta $ 
respectively, $\hat {n}_{i\sigma }^\gamma = \gamma _{i\sigma }^ + \gamma 
_{i\sigma } $ is the electron number operator for electrons with spin 
$\sigma $ on the sub-lattice $\gamma = \alpha ,\beta $, $t_{eff} = t \cdot 
b^{AF}$ is the effective hopping integral.

Now we will derive the equations for dispersion relation, particle number 
and magnetization in the antiferromagnetic state using Hamiltonian (\ref{eq44}) and 
the Green function technique. For the on-site interaction, $U$, we use the 
standard CPA approximation, which for simplicity, in further analysis, is 
treated in the weak and strong correlation limits. For the weak correlation 
we use a first order approximation in interaction constant over the 
bandwidth. This is equivalent to the Hartree-Fock approximation. The second 
case is the high correlation approximation, $U > > D$, , which will be easy 
to extend for the arbitrary strength of the on-site interaction later on.

The main idea of the CPA formalism \cite{8} is used now. We split the above 
stochastic Hamiltonian (\ref{eq44}) into a homogeneous part 

\begin{equation}
\label{eq45}
H_0 = - t_{eff} \sum\limits_{i\sigma } {\left( {\alpha _{i\sigma }^ + \beta 
_{i\sigma } + \beta _{i\sigma }^ + \alpha _{i\sigma } } \right)} - \mu 
\sum\limits_{i\sigma } {(\hat {n}_{i\sigma }^\alpha + \hat {n}_{i\sigma 
}^\beta )} + \sum\limits_{i\sigma } {\Sigma _\sigma ^\alpha \,\hat 
{n}_{i\sigma }^\alpha } + \sum\limits_{i\sigma } {\,\Sigma _\sigma ^\beta 
\hat {n}_{i\sigma }^\beta } \quad ,
\end{equation}
and a stochastic part 

\begin{equation}
\label{eq46}
H_I = \sum\limits_{i\sigma } {(\tilde {V}_\sigma ^\alpha - \Sigma _\sigma 
^\alpha )\,\hat {n}_{i\sigma }^\alpha } + \sum\limits_{i\sigma } {\,(\tilde 
{V}_\sigma ^\beta - \Sigma _\sigma ^\beta )\hat {n}_{i\sigma }^\beta } \quad ,
\end{equation}
where

\begin{equation}
\label{eq47}
\mu = \mu _0 - zVn - 2z\Delta tI_{AF} 
\end{equation}
is the effective chemical potential, $\Sigma _\sigma ^\gamma $ are the self 
energies on sites $\gamma = \alpha ,\beta $ for electrons with spin $\sigma 
$ and $\tilde {V}_\sigma ^{\alpha (\beta )} $ are the stochastic potentials 
given by 

\begin{equation}
\label{eq47a}
\tilde {V}_\sigma ^{\alpha (\beta )} = \left\{ {\begin{array}{c}
 \tilde {V}_{1\sigma }^{\alpha (\beta )} = - Fn_\sigma ^{\alpha (\beta )} - 
z\left( {J + 2t_{ex} I_{AF} } \right)n_\sigma ^{\beta (\alpha )} \\ 
 \tilde {V}_{2\sigma }^{\alpha (\beta )} = U - Fn_\sigma ^{\alpha (\beta )} 
- z\left( {J + 2t_{ex} I_{AF} } \right)n_\sigma ^{\beta (\alpha )} \\ 
 \end{array}} \textrm{, with probabilities} \begin{array}{c}
 P_{1\sigma }^{\alpha (\beta )} = 1 - n_{ - \sigma }^{\alpha (\beta )} \\ 
 P_{2\sigma }^{\alpha (\beta )} = n_{-\sigma }^{\alpha (\beta )} \\ 
 \end{array}
 \right . 
\end{equation}

The self-energies $\Sigma _\sigma ^\gamma $ fulfill the CPA equations

\be
\sum\limits_{i = 1}^2 {P_{i\sigma }^\gamma \frac{\tilde {V}_{i\sigma 
}^\gamma - \Sigma _\sigma ^\gamma }{1 - \left( {\tilde {V}_{i\sigma }^\gamma 
- \Sigma _\sigma ^\gamma } \right)F_\sigma ^\gamma (\varepsilon )}} = 0 
\quad \textrm{, for} \quad \gamma = \alpha (\beta ) ,
\label{eq47b}
\ee 
with the Slater-Koster function $F_\sigma ^\gamma (\varepsilon )$ for the 
sub-lattice $\gamma $ in the following form

\begin{equation}
\label{eq48}
F_\sigma ^\gamma (\varepsilon ) = \frac{1}{N}\sum\limits_k {G_\sigma 
^{\gamma \gamma } (\varepsilon ,k)} \quad ,
\end{equation}
and $G_\sigma ^{\gamma \gamma } (\varepsilon ,k)$ given by Eq. (\ref{eq51}) below.

After dropping the paramagnetic part the stochastic potential (48) takes on 
the following form 

\be
\tilde {V}_\sigma ^{\alpha (\beta )} = \left\{ {\begin{array}{c}
 \tilde {V}_{1\sigma }^{\alpha (\beta )} = \mp \Delta \\ 
 \tilde {V}_{2\sigma }^{\alpha (\beta )} = U \mp \Delta \\ 
 \end{array}} \right. \quad \textrm{,    with} \quad \begin{array}{c}
 P_{1\sigma }^{\alpha (\beta )} = 1 - n_{ \mp \sigma } \\ 
 P_{2\sigma }^{\alpha (\beta )} = n_{ \mp \sigma } \\ 
 \end{array} , 
 \label{eq51a}
\ee
where 

\be
n_{\pm \sigma } = \frac{n\pm m}{2} \quad ,
\quad
\Delta = F_{tot}^{AF} \frac{m}{2} \quad \textrm{, and} \quad F_{tot}^{AF} = F - z\left( {J 
+ 2t_{ex} I_{AF} } \right) . 
\label{eq52a}
\ee

This simplification of the Eq. (49) allows to see that the self-energies 
fulfill the relations; $\Sigma _\sigma ^\alpha = \Sigma _{ - \sigma }^\beta 
\equiv \Sigma _ + $, and $\Sigma _{ - \sigma }^\alpha = \Sigma _\sigma 
^\beta \equiv \Sigma _ - $. Therefore the solution of the problem can be 
reduced to finding only the self-energies, $\Sigma _\pm $. The 
interpretation of the parameter $\Delta $ as the antiferromagnetic energy 
gap will be given below.

We transform Hamiltonian (\ref{eq45}) into the momentum space, and use it in the 
equations of motion for the Green functions 

\begin{equation}
\label{eq49}
\varepsilon \,\left\langle {\left\langle {A;B} \right\rangle } \right\rangle 
_\varepsilon = \left\langle {\left[ {A,B} \right]_ + } \right\rangle + 
\left\langle {\left\langle {\left[ {A,H_0 } \right]_{\, - } ;B} 
\right\rangle } \right\rangle _\varepsilon \quad ,
\end{equation}
where $(A,B) \in \left( {\alpha _{k\sigma }^ + ,\alpha _{k\sigma } ,\beta 
_{k\sigma }^ + ,\beta _{k\sigma } } \right)$. As a result the following 
equations are obtained

\begin{equation}
\label{eq50}
\left[ {\begin{array}{*{20}c}
   {\varepsilon  + \mu  - \Sigma _\sigma ^\alpha  } & { - \varepsilon _k }  \\
   { - \varepsilon _k } & {\varepsilon  + \mu  - \Sigma _\sigma ^\beta  }  \\
\end{array}} \right]\left[ {\begin{array}{*{20}c}
   {G_\sigma ^{\alpha \alpha } (\varepsilon ,k)} & {G_\sigma ^{\alpha \beta } (\varepsilon ,k)}  \\
   {G_\sigma ^{\beta \alpha } (\varepsilon ,k)} & {G_\sigma ^{\beta \beta } (\varepsilon ,k)}  \\
\end{array}} \right] = \left[ {\begin{array}{*{20}c}
   1 & 0  \\
   0 & 1  \\
\end{array}} \right] ,
\end{equation}
with e.g. $G_\sigma ^{\alpha \alpha } (\varepsilon ,k) = < < \alpha 
_{k\sigma } ;\alpha _{k\sigma }^ + > > $, $G_\sigma ^{\beta \alpha } 
(\varepsilon ,k) = < < \beta _{k\sigma } ;\alpha _{k\sigma }^ + > > $.

Solving this set of equations we arrive at the following expressions for the 
Green functions 

\begin{equation}
\label{eq51}
G_\sigma ^{\alpha \alpha (\beta \beta )} (\varepsilon ,k) = \frac{1}{2}\sqrt 
{\frac{\varepsilon + \mu - \Sigma _\sigma ^{\beta (\alpha )} }{\varepsilon + 
\mu - \Sigma _\sigma ^{\alpha (\beta )} }} \left[ {G\left( {\varepsilon 
_{eff} ,k} \right) - G\left( { - \varepsilon _{eff} ,k} \right)} \right] 
\quad ,
\end{equation}
with

\begin{equation}
\label{eq52}
\varepsilon _{eff} = \sqrt {\left( {\varepsilon + \mu - \Sigma _ + } 
\right)\left( {\varepsilon + \mu - \Sigma _ - } \right)} \quad ,
\end{equation}
and $G(\varepsilon _{eff} ,k)$ given by

\begin{equation}
\label{eq53}
G\left( {\varepsilon _{eff} ,k} \right) = \frac{1}{\varepsilon _{eff} - 
\varepsilon _k } \quad ,
\quad
\varepsilon _k = \varepsilon _k^0 b^{AF} \quad .
\end{equation}

Remembering that 

\begin{equation}
\label{eq54}
\rho _\sigma ^\gamma \left( \varepsilon \right) = - \frac{1}{\pi } \rm Im \left[ 
{F_\sigma ^\gamma \left( \varepsilon \right)} \right]
\end{equation}
we obtain

\begin{equation}
\label{eq55}
\rho _\sigma ^{\alpha (\beta )} = - \frac{1}{\pi } \rm Im \sqrt {\frac{\varepsilon 
+ \mu - \Sigma _\sigma ^{\beta (\alpha )} }{\varepsilon + \mu - \Sigma 
_\sigma ^{\alpha (\beta )} }} \frac{1}{b^{AF}}F_0 \left( {\frac{\varepsilon 
_{eff} }{b^{AF}}} \right) \equiv - \frac{1}{\pi } \rm Im F^\pm \quad ,
\end{equation}
where $F_0 (\varepsilon )$ is the unperturbed Slater-Koster function given 
by Eq. (\ref{eq32}).

Using Eq. (\ref{eq48}) we obtain the expressions for electron numbers

\begin{equation}
\label{eq56}
n_\sigma ^\gamma = \int\limits_{ - \infty }^\infty {\frac{1}{1 + 
e^{\varepsilon \mathord{\left/ {\vphantom {\varepsilon {kT}}} \right. 
\kern-\nulldelimiterspace} {kT}}}\left( { - \frac{1}{\pi }} \right)} 
\rm Im \left[ {F_\sigma ^\gamma \left( \varepsilon \right)} \right]d\varepsilon = 
\frac{1}{N}\sum\limits_k {\int\limits_{ - \infty }^\infty {\frac{1}{1 + 
e^{\varepsilon \mathord{\left/ {\vphantom {\varepsilon {kT}}} \right. 
\kern-\nulldelimiterspace} {kT}}}\left( { - \frac{1}{\pi }} \right)} 
\rm Im \left[ {G_\sigma ^{\gamma \gamma } \left( {\varepsilon ,k} \right)} 
\right]} d\varepsilon \quad .
\end{equation}

Treating the self-energies $\Sigma _\sigma ^\gamma $ as the solutions of Eq. 
(49), and using it only in the first order approximation with respect to the 
Coulomb on-site repulsion, $U$, we obtain 

\begin{equation}
\label{eq57}
\Sigma _{\pm \sigma }^\alpha = \Sigma _\pm \cong \sum\limits_{i = 1}^2 
{P_{i\pm \sigma }^\alpha \tilde {V}_{i\pm \sigma }^\alpha } = \left[ {U - F 
- z\left( {J + 2t_{ex} I_{AF} } \right)} \right]\frac{n}{2} \mp \Delta \quad ,
\end{equation}
where the antiferromagnetic energy gap $\Delta $ is given by the expression 

\begin{equation}
\label{eq58}
\Delta = \left[ {U + F_{tot}^{AF} } \right]\frac{m}{2} \quad .
\end{equation}

This expression is different from Eq. (52) by adding to the effective field 
the Coulomb repulsion, $U$, which in the case of the weak interaction is 
also treated in the Hartree-Fock approximation.

Neglecting the paramagnetic part in self-energies we have 

\begin{equation}
\label{eq59}
\Sigma _\pm = \mp \Delta .
\end{equation}

Inserting these self-energies into Eq. (\ref{eq51}) and redefining $\varepsilon + 
\mu $ as $\varepsilon $ we obtain

\begin{equation}
\label{eq60}
G_\sigma ^{\alpha \alpha \left( {\beta \beta } \right)} (\varepsilon ,k) = 
\frac{1}{2}\sqrt {\frac{\varepsilon \mp \Delta }{\varepsilon \pm \Delta }} 
\left[ {G\left( {\varepsilon _{eff} ,k} \right) - G\left( { - \varepsilon 
_{eff} ,k} \right)} \right] \quad .
\end{equation}

From this equation using Eqs (\ref{eq48}) and (\ref{eq54}) one can obtain the following 
relation for the DOS on sites $\alpha ,\beta $

\begin{equation}
\label{eq61}
\rho _\sigma ^{\alpha \left( \beta \right)} \left( \varepsilon \right) = 
\sqrt {\frac{\varepsilon \mp \Delta }{\varepsilon \pm \Delta }} \rho \left( 
{\varepsilon _{eff} } \right) \equiv \rho _\pm (\varepsilon ) \quad .
\end{equation}

In this form it is evident that the AF DOS will vanish between energies $ - 
\Delta $ and $ + \Delta $ (zero energy was assumed in the center of the 
atomic band or approximately at the atomic level). Hence, the quantity 
$\Delta $ is the AF energy gap. The schematic shape of the DOS is shown in 
Fig. 1.

\begin{figure}[htbp]
\begin{center}
\epsfxsize8cm
\epsffile{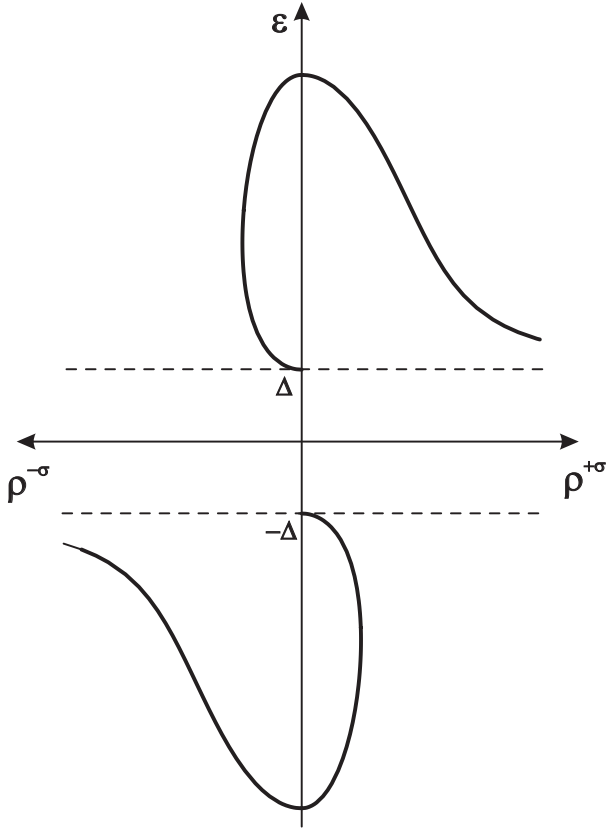}
\caption{Schematic DOS in the antiferromagnetic state. The energy gap extends 
from $ - \Delta $ to $ + \Delta $.}
\label{fig1}
\end{center}
\end{figure}

One can be prove that if the old distribution is normalized to unity; 

\begin{equation}
\label{eq62}
\int\limits_{ - D}^D {\rho (\varepsilon ^0)d\varepsilon ^0 = 1} \quad ,
\end{equation}
then the new one given by Eq. (\ref{eq61}) is also normalized. 

Since $\rho _ + (\varepsilon ) = \rho _ - ( - \varepsilon )$, one can write 
that

\begin{equation}
\label{eq63}
\int\limits_{ - \sqrt {D_{eff}^2 + \Delta ^2} }^{\sqrt {D_{eff}^2 + \Delta 
^2} } {\rho _ + (\varepsilon )d\varepsilon } = \int\limits_{ - \sqrt 
{D_{eff}^2 + \Delta ^2} }^{\sqrt {D_{eff}^2 + \Delta ^2} } 
{\frac{1}{2}\left[ {\rho _ + (\varepsilon ) + \rho _ - ( - \varepsilon )} 
\right]d\varepsilon } = \frac{1}{2}\int\limits_{ - \sqrt {D_{eff}^2 + \Delta 
^2} }^{\sqrt {D_{eff}^2 + \Delta ^2} } {\left[ {\rho _ + (\varepsilon ) + 
\rho _ - (\varepsilon )} \right]d\varepsilon } \quad ,
\end{equation}
where $D_{eff} = zt \cdot b^{AF}$. Inserting

\begin{equation}
\label{eq64}
\rho _ + (\varepsilon ) + \rho _ - (\varepsilon ) = \left( {\sqrt 
{\frac{\varepsilon - \Delta }{\varepsilon + \Delta }} + \sqrt 
{\frac{\varepsilon + \Delta }{\varepsilon - \Delta }} } \right)\rho \left( 
{\varepsilon _{eff} } \right) \equiv \frac{2\varepsilon }{\varepsilon _{eff} 
}\rho \left( {\varepsilon _{eff} } \right)
\end{equation}
and using the identity obtained from Eq. (\ref{eq52}); $d\varepsilon = 
\frac{\varepsilon _{eff} }{\varepsilon }d\varepsilon _{eff} $, one obtains 
from Eq. (\ref{eq63})

\begin{equation}
\label{eq65}
\int\limits_{ - \sqrt {D_{eff}^2 + \Delta ^2} }^{\sqrt {D_{eff}^2 + \Delta 
^2} } {\rho _ + (\varepsilon )d\varepsilon } = \int\limits_{ - D_{eff} 
}^{D_{eff} } {\rho (\varepsilon _{eff} )d\varepsilon _{eff} } = 1 \quad .
\end{equation}

Restricting as before only to the first order approximation for the 
self-energies (Eq. (\ref{eq59})) in Eq. (\ref{eq56}) we obtain the following expressions for 
electron numbers

\begin{equation}
\label{eq66}
n_{\pm \sigma }^\alpha = n_{ \mp \sigma }^\beta = \frac{1}{2N}\sum\limits_k 
{\left[ {P_k^{\pm \sigma } f(E_k ) + P_k^{ \mp \sigma } f( - E_k )} \right]} 
\quad ,
\end{equation}
where $f(E_k )$ is the Fermi function

\begin{equation}
\label{eq67}
f(E_k ) = \frac{1}{1 + e^{{\left( {E_k - \mu } \right)} \mathord{\left/ 
{\vphantom {{\left( {E_k - \mu } \right)} {kT}}} \right. 
\kern-\nulldelimiterspace} {kT}}} \quad ,
\end{equation}

\begin{equation}
\label{eq68}
E_k = \sqrt {\varepsilon _k^2 + \Delta ^2} \quad ,
\quad
S_k^{\pm \sigma } = \sqrt {\frac{E_k \mp \,\Delta }{E_k \pm \,\Delta }} 
\quad ,
\end{equation}
and

\begin{equation}
\label{eq69}
P_k^{\pm \sigma } = \frac{\varepsilon _k S_k^{\pm \sigma } }{E_k }
\end{equation}
is the occupation probability of state $(k,\pm \sigma )$.

The expression for electron occupation in the antiferromagnetic state in the 
presence of external field $H$ will be different from Eq. (\ref{eq36}), which was 
suitable for ferromagnetism. Now, the external field will act differently on 
both spins $\pm \sigma $ in two sub-lattices $\alpha $ and$\beta $. 
Therefore its energy has to be included into the energy of the internal 
exchange field in the process of diagonalization. It will result in the 
following formula for the electron occupation on the sub-lattice $\alpha $

\be
n_\sigma ^\alpha = \int\limits_{ - \infty }^\infty {\rho _\sigma ^\alpha 
\left( {\varepsilon ,x^\sigma } \right)\frac{d\varepsilon }{1 + e^{{\left( 
{\varepsilon - \mu } \right)} \mathord{\left/ {\vphantom {{\left( 
{\varepsilon - \mu } \right)} {kT}}} \right. \kern-\nulldelimiterspace} 
{kT}}}} \quad \textrm{, where} \quad x^\sigma = \sigma \Delta + \sigma \mu _B H . 
\label{eq74a}
\ee

The precise form of the density of states $\rho _\sigma ^{\alpha \left( 
\beta \right)} \left( \varepsilon \right)$ comes from Eq. (\ref{eq55}) (or (\ref{eq61}) in 
the case of the Hartree-Fock approximation). As it can be seen, it depends 
on the initial DOS and on the result of the diagonalization of two 
sub-lattices Hamiltonian.

\vskip1cm 

\noindent{\Large{\bf 3. FREE ENERGY, STATIC MAGNETIC SUSCEPTIBILITY}}

\vskip0.5cm 

For a simple magnetic material, the free energy can be written 

\begin{equation}
\label{eq70}
F_m = F_0 + a_2 m^2 + a_4 m^4 + ... = F_0 + \mu _B^2 m^2 / \chi + a_4 m^4 + 
...
\end{equation}
with $\chi $ being the magnetic susceptibility.

The electron occupation of the {\bf ferromagnetic} state in the presence of 
an external magnetic field $H$ is given by the expression (\ref{eq36}) with added 
shift from the magnetic field

\begin{equation}
\label{eq71}
n_\sigma = \int\limits_{ - \infty }^\infty {\rho _{^\sigma } \left( 
\varepsilon \right)\frac{d\varepsilon }{1 + e^{{\left( {\varepsilon - \mu + 
M^\sigma - \sigma \mu _B H} \right)} \mathord{\left/ {\vphantom {{\left( 
{\varepsilon - \mu + M^\sigma - \sigma \mu _B H} \right)} {kT}}} \right. 
\kern-\nulldelimiterspace} {kT}}}} \quad ,
\end{equation}
where $\mu _B $ is the Bohr magneton.

The magnetic susceptibility is given by 

\begin{equation}
\label{eq72}
\chi = \mu _B \left( {\frac{\partial \left( {n_\sigma - n_{ - \sigma } } 
\right)}{\partial H}} \right)_{H = 0} \quad .
\end{equation}

As a result of inserting Eq. (\ref{eq71}) into it we arrive at the following 
equation for the static ferromagnetic susceptibility

\begin{equation}
\label{eq73}
\chi = \frac{2\mu _B^2 I_T }{1 - K - F_{tot}^F I_T } \quad ,
\end{equation}
where $K$ is the correlation factor defined mathematically by the following 
equation

\begin{equation}
\label{eq74}
\begin{array}{c}
 K = \int\limits_{ - \infty }^\infty {\frac{\partial \rho _\sigma \left( 
\varepsilon \right)}{\partial m}\frac{d\varepsilon }{1 + e^{{\left( 
{\varepsilon - \mu + M^\sigma } \right)} \mathord{\left/ {\vphantom {{\left( 
{\varepsilon - \mu + M^\sigma } \right)} {kT}}} \right. 
\kern-\nulldelimiterspace} {kT}}}} - \int\limits_{ - \infty }^\infty 
{\frac{\partial \rho _{ - \sigma } \left( \varepsilon \right)}{\partial 
m}\frac{d\varepsilon }{1 + e^{{\left( {\varepsilon - \mu + M^{ - \sigma }} 
\right)} \mathord{\left/ {\vphantom {{\left( {\varepsilon - \mu + M^{ - 
\sigma }} \right)} {kT}}} \right. \kern-\nulldelimiterspace} {kT}}}} \\ 
 \mbox{ } = 2\int\limits_{ - \infty }^\infty {\frac{\partial \rho _\sigma 
\left( \varepsilon \right)}{\partial m}\frac{d\varepsilon }{1 + e^{{\left( 
{\varepsilon - \mu + M^\sigma } \right)} \mathord{\left/ {\vphantom {{\left( 
{\varepsilon - \mu + M^\sigma } \right)} {kT}}} \right. 
\kern-\nulldelimiterspace} {kT}}}} \\ 
 \end{array} \quad ,
\end{equation}
$F_{tot}^F $is the total Stoner exchange field, which on the base of Eq. (\ref{eq25}) 
can be written as

\begin{equation}
\label{eq75}
F_{tot}^F = - \left( {\left. {\frac{\partial M^\sigma }{\partial m}} 
\right|_{m \to 0} - \left. {\frac{\partial M^{ - \sigma }}{\partial m}} 
\right|_{m \to 0} } \right) = - 2\left. {\frac{\partial M^\sigma }{\partial 
m}} \right|_{m \to 0} ,
\end{equation}
and

\begin{equation}
\label{eq76}
I_T = \int\limits_{ - \infty }^\infty {\rho _{m = 0} \left( \varepsilon 
\right)P_T \left( \varepsilon \right)d\varepsilon } ,
\quad
P_T \left( \varepsilon \right) = - \frac{\partial f\left( \varepsilon 
\right)}{\partial \varepsilon } = f^2\left( \varepsilon \right)e^{{\left( 
{\varepsilon - \mu } \right)} \mathord{\left/ {\vphantom {{\left( 
{\varepsilon - \mu } \right)} {kT}}} \right. \kern-\nulldelimiterspace} 
{kT}}\frac{1}{kT} \quad ,
\end{equation}
$f\left( \varepsilon \right)$ is a Fermi function.

Returning with the expression (\ref{eq73}) to Eq. (\ref{eq70}) we arrive at a new form of 
Landau expansion which is

\begin{equation}
\label{eq77}
F_m = F_0 + m^2\frac{1 - K - F_{tot}^F I_T }{2I_T } + a_4 m^4 + ... \quad .
\end{equation}

The critical (minimum) value of the total Stoner field creating 
ferromagnetism; $(F_{tot}^F )_{\min } = F_{tot}^{cr} $, is obtained from the 
zero of susceptibility denominator (Eq. (\ref{eq73})) or zero of the numerator in 
Eq. (\ref{eq77}), around which changes the sign of the second order term. Hence, we 
arrive at the well-known Stoner criterion (see \cite{56,57})

\begin{equation}
\label{eq78}
F_{tot}^{cr} = \frac{1 - K}{I_T } \quad ,
\end{equation}
which is modified here by including the correlation factor $K$.

Using the Stoner condition$F_{tot}^{cr} = {\left( {1 - K} \right)} 
\mathord{\left/ {\vphantom {{\left( {1 - K} \right)} {I_T }}} \right. 
\kern-\nulldelimiterspace} {I_T }$, back in the expression for 
susceptibility, Eq. (\ref{eq73}), and in Landau expansion (\ref{eq77}) we arrive at their 
corrected forms

\begin{equation}
\label{eq79}
\chi = \frac{2\mu _B^2 }{F_{tot}^{cr} - F_{tot}} \quad ,
\end{equation}

\begin{equation}
\label{eq80}
F_m = F_0 + \frac{m^2}{2}\left( {F_{tot}^{cr} - F_{tot} } \right) + a_4 m^4 
+ ... \quad ,
\end{equation}
since $a_4 > 0$, the existence of the minimum in the above equation for 
$\mbox{F}_m $ will depend on the sign of $a_2 = \frac{1}{2}(F_{tot}^{cr} - 
F_{tot} )$ or equivalently on the sign of the difference: $F_{tot}^{cr} - 
F_{tot} $. As a result, the simple Stoner criterion (Eq. (\ref{eq78})) will lead to 
the ferromagnetic instability with nonzero magnetic moment being given by 
the minimum of $\mbox{F}_m $ curve in Fig. 2. This minimum appears only when 
existing in a given material, $F_{tot}^F > F_{tot}^{cr} $, the $F_{tot}^{cr} 
$ being given by condition of Eq. (\ref{eq78}). It is not mistake that in Eqs (\ref{eq79}) 
and (\ref{eq80}) we have $F_{tot} $ without the upper-script $F$, since as we will 
see below those two equations are valid without any changes for both F and 
AF ordering.

To calculate the static {\bf antiferromagnetic} susceptibility we will apply 
Eq. (\ref{eq72}) to the susceptibility of the sub-lattice \footnote{This susceptibility is a susceptibility of one 
sub-lattice. To find its relation with the total experimental susceptibility 
one has to take into consideration the direction of the external field with 
respect to the easy magnetic axis (see e.g. \cite{58}, \cite{59})}

\begin{equation}
\label{eq81}
\chi = \mu_B \left( {\frac{\partial \left( {n_\sigma ^\alpha - n_{ - \sigma 
}^\alpha } \right)}{\partial H}} \right)_{H = 0} \quad ,
\end{equation}
and we will use expression (74) for the electron occupation on sub-lattice 
$\alpha $ in the presence of an external field, arriving at the following 
equation for the static antiferromagnetic susceptibility

\begin{equation}
\label{eq82}
\chi = \frac{2\mu _B^2 K_x }{1 - K_x F_{tot}^{AF} } \quad ,
\end{equation}
where the correlation factor $K_x $ in the antiferromagnetic case can be 
written as

\begin{equation}
\label{eq83}
K_x = \frac{1}{2}\left( {\frac{\partial n_\sigma ^\alpha }{\partial 
x^\sigma } - \frac{\partial n_{ - \sigma }^\alpha }{\partial x^{ - \sigma 
}}} \right) = \frac{1}{2}\int\limits_{ - \infty }^\infty {\left( {\frac{\partial \rho 
_\sigma ^\alpha \left( \varepsilon \right)}{\partial x^\sigma } - 
\frac{\partial \rho _{ - \sigma }^\alpha \left( \varepsilon 
\right)}{\partial x^{ - \sigma }}} \right)\frac{d\varepsilon }{1 + 
e^{{\left( {\varepsilon - \mu } \right)} \mathord{\left/ {\vphantom {{\left( 
{\varepsilon - \mu } \right)} {kT}}} \right. \kern-\nulldelimiterspace} 
{kT}}}} = \int\limits_{ - \infty }^\infty {\frac{\partial \rho _\sigma 
^\alpha \left( \varepsilon \right)}{\partial x^\sigma }} f\left( \varepsilon 
\right)d\varepsilon  
.
\end{equation}

Using the condition for the zero of susceptibility denominator we arrive at 
the critical value of the total field

\begin{equation}
\label{eq84}
F_{tot}^{cr} = \frac{1}{K_x \left( n \right)} \quad .
\end{equation}

After inserting this value into the sub-lattice susceptibility (Eq. (\ref{eq82})) 
and next to the formula for Landau expansion (Eq. (\ref{eq70})) we obtain again the 
same form of the susceptibility as for ferromagnetism (Eq. (\ref{eq79})), and of the 
free energy expansion as given by Eq. (\ref{eq80}) for ferromagnetism. Thus the 
susceptibility of Eq. (\ref{eq79}) and the energy expansion of Eq. (\ref{eq80}) are the same 
for F and AF. The difference between these two orderings lies in the 
different values of critical (minimum) field. For F this field is given by 
Eq. (\ref{eq78}) and for AF by Eq. (\ref{eq84}). In both cases when existing in a given 
material$F_{tot}^{F\left( {AF} \right)} > F_{tot}^{cr} $, then the second 
order term in energy expansion vs. magnetization becomes negative and we 
have a nonzero equilibrium value of the ordering parameter $m$, see Fig. 2.

\begin{figure}[htbp]
\begin{center}
\epsfxsize11cm
\epsffile{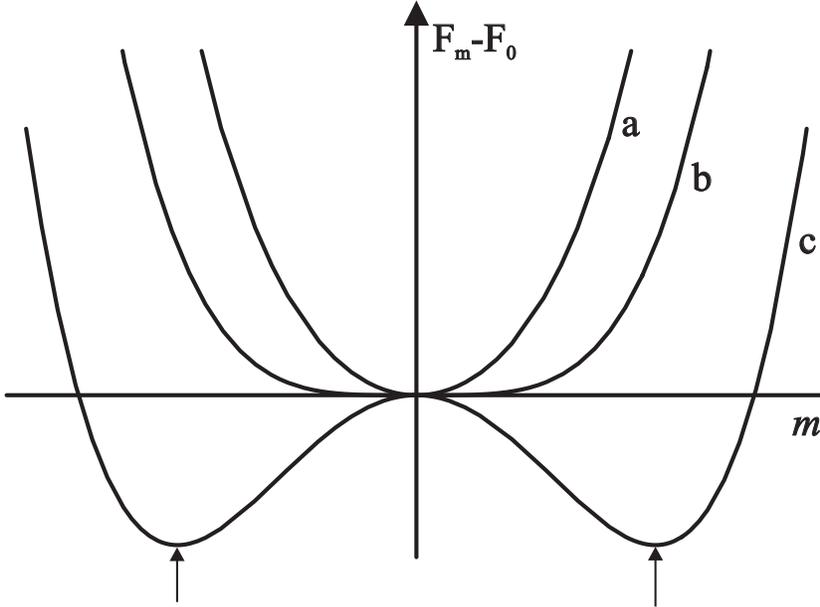}
\caption{ Dependence of Landau free energy on the magnetization $m$ for various 
values of $a_2 = \frac{1}{2}\left( {F_{tot}^{cr} - F_{tot}^{F\left( {AF} 
\right)} } \right)$. Curve (a)- $F_{tot}^{F\left( {AF} \right)} < 
F_{tot}^{cr} $, curve (b)- $F_{tot}^{F\left( {AF} \right)} = F_{tot}^{cr} $, 
curve (c)- $F_{tot}^{F\left( {AF} \right)} > F_{tot}^{cr} $.
}
\label{fig2}
\end{center}
\end{figure}

\vskip1cm 

\noindent{\Large{\bf 4 FERROMAGNETISM}}

\vskip0.5cm 

{\bf 4.1 ONSET OF FERROMAGNETISM}

\vskip0.5cm 

To compute the ferromagnetic criterion from Eq. (\ref{eq78}) we have to calculate 
the correlation factor, $K$, defined in Eq. (\ref{eq74}). We will calculate it now 
in the CPA approximation, assuming that we know the self-energy $\Sigma _\sigma$. The total change of the DOS with 
magnetization, $K$, which 
enhances the possibility of creating magnetic ordering, is given from Eq. 
(\ref{eq74}) as

\begin{equation}
\label{eq85}
K = K_U + K_b \quad ,
\end{equation}
where

\begin{equation}
\label{eq86}
K_U = 2\int\limits_{ - \infty }^{ + \infty } {\frac{\partial \rho _\sigma 
(\varepsilon )}{\partial \Sigma _\sigma }\frac{\partial \Sigma _\sigma 
}{\partial m}} f(\varepsilon )d\varepsilon 
\end{equation}
and

\begin{equation}
\label{eq87}
K_b = 2\int\limits_{ - \infty }^{ + \infty } {\frac{\partial \rho _\sigma 
(\varepsilon )}{\partial b^\sigma }\frac{\partial b^\sigma }{\partial m}} 
f(\varepsilon )d\varepsilon \quad .
\end{equation}

Factor $K_U $ describes the role of on-site interaction $U$, and $K_b $ the 
role of inter-site interactions in creating magnetization. 

The schematic depiction of the DOS deformed by the on-site correlation $U$is 
shown in Fig. 3. 

\begin{figure}[htbp]
\begin{center}
\epsfxsize8cm
\epsffile{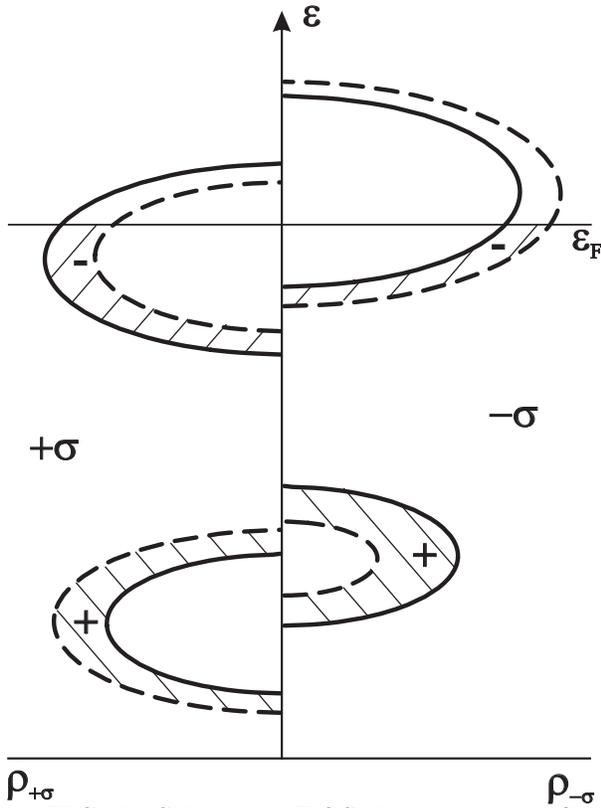}
\caption{Schematic DOS showing the influence of the strong on-site Coulomb 
correlation, $U$. The paramagnetic DOS for both spins, $\pm \sigma $, are 
solid lines. At $U$ which is strong enough to split the band into two 
sub-bands, lower sub-bands have the capacity of $1 - n^{ - \sigma }$ for $ + 
\sigma $ electrons, and $1 - n^\sigma $ for $ - \sigma $ electrons. The 
changes in the spin electron densities integrated over energy are the shaded 
areas in this figure, and they are equal to the correlation factor $K_U $. 
The shift between $ + \sigma $ and $ - \sigma $ electrons is created by the 
assumed exchange field.
}
\label{fig3}
\end{center}
\end{figure}

The schematic depiction of the DOS deformed by the inter-site correlation is 
shown in Fig. 4.

\begin{figure}[htbp]
\begin{center}
\epsfxsize8cm
\epsffile{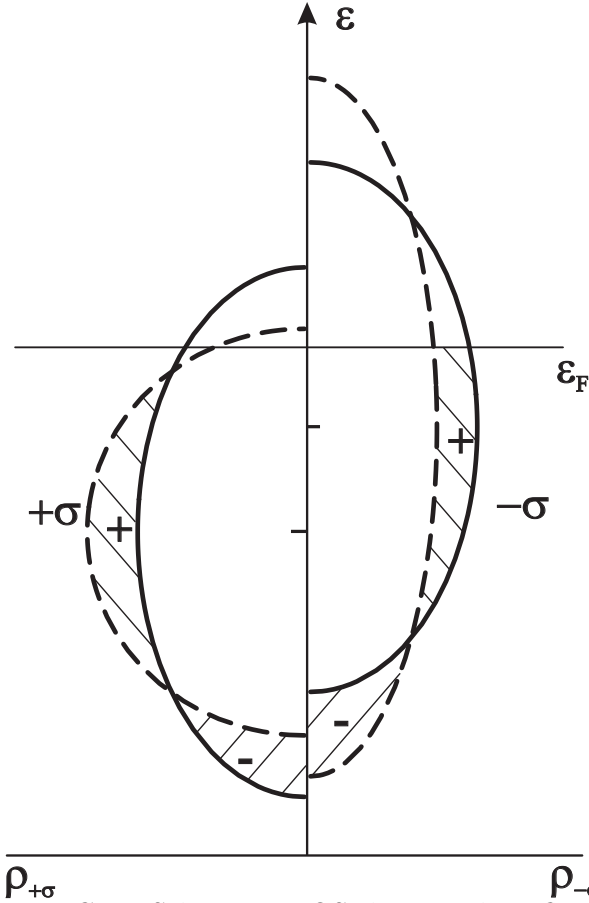}
\caption{Schematic DOS showing the influence of the inter-site interactions. 
The paramagnetic DOS for both spins, $\pm \sigma $, are solid lines. The 
inter-site interactions change the relative width of the bands with respect 
to each other (described by the $b^\sigma $ factors, see Eq. (\ref{eq22})). The 
Stoner field, which would displace the bands with respect to each other, is 
assumed to be nonzero. The shaded areas in this figure are the correlation 
factor $K_b $.}
\label{fig4}
\end{center}
\end{figure}

Comparing Figs 3 and 4 one can see that the on-site correlation causes an 
increase in capacity of the majority spin band (or lower sub-band in the 
case of split band), while the inter-site correlation causes a decrease in 
width of the majority spin band. There is an important difference between 
these two effects. The situation from Fig. 3 can not lead by itself, without 
the exchange field, to ferromagnetism (see \cite{9,10}), while the majority spin 
bandwidth decrease (see Fig 4) can. However, there is the problem with the 
magnitude of the interaction constants, which can lead to this ordering. 
There is also a difference in the order of approximation necessary to obtain 
different correlations. The band shape change shown in the Fig. 3 can be 
obtained only in the approximations higher than the 1$^{st}$ order in $U / 
D$, while the change shown in Fig. 4 is obtained already in the 1$^{st}$ 
order approximation (Hartree-Fock). 

After inserting Eq. (\ref{eq85}) into Eq. (\ref{eq73}) one can write for the static 
ferromagnetic susceptibility

\begin{equation}
\label{eq88}
\chi = \frac{2\mu _B^2 I_T }{1 - K_U - K_b - F_{tot}^F I_T } \quad .
\end{equation}

At zero temperature we obtain from Eq. (\ref{eq76}) that; $I_T = {\rho 
^0(\varepsilon _F^0 )} \mathord{\left/ {\vphantom {{\rho ^0(\varepsilon _F^0 
)} {b^0}}} \right. \kern-\nulldelimiterspace} {b^0}$, where $\varepsilon 
_F^0 $ is the Fermi level for the system without inter-site interactions and 
$b^0$ is the bandwidth change parameter in the paramagnetic state due to the 
inter-site interactions. This will modify the last relations into the 
following form

\begin{equation}
\label{eq89}
\chi = \frac{2\mu _B^2 \rho ^0(\varepsilon _F^0 )}{\left( {1 - K_U - K_b } 
\right)b^0 - F_{tot}^F \rho ^0(\varepsilon _F^0 )} \quad .
\end{equation}

Comparing the above equation with the equation for susceptibility when there 
is only the Hartree-Fock exchange field

\begin{equation}
\label{eq90}
\chi = \frac{2\mu _B^2 \rho ^0(\varepsilon _F^0 )}{1 - F_{tot}^F \rho 
^0(\varepsilon _F^0 )}
\end{equation}
one can see that the ferromagnetic state will be favored by both positive 
correlation factors: $K_U $ and $K_b $, and by a decrease in the bandwidth 
$b^0 < 1$ due to the inter-site interactions.

From the zero of the susceptibility denominator (Eq. (\ref{eq89})) and Eqs (\ref{eq75}), 
(\ref{eq23}), one can calculate the critical on-site exchange interaction.

\begin{equation}
\label{eq91}
F^{cr} = \frac{\left( {1 - K_U - K_b } \right)b^0}{\rho (\varepsilon _F )} - 
zJ - 4D\frac{\partial }{\partial m}\left[ {\frac{\left( {1 - S} 
\right)^2}{2}I_{ - \sigma } n_\sigma - I_{ - \sigma } \left( {1 - S} 
\right)} \right] \quad ,
\end{equation}
where the kinetic interactions were expressed by the common factor $S$, see 
Eq. (\ref{eq5}). 

The inter-site correlation factor given by Eq. (\ref{eq87}) can be written as 

\be
K_b = - 2\int\limits_{ - D}^D {\rho _{m = 0} \left( \varepsilon \right)} P_T 
\left( \varepsilon \right)\varepsilon \frac{\partial b^\sigma }{\partial 
m}d\varepsilon  . 
\label{eq97a}
\ee

At zero temperature we have $P_T \left( \varepsilon \right) \Rightarrow 
\frac{\delta \left( {\varepsilon - \varepsilon _F^0 } \right)}{b^0}$, which 
produces the following expression

\begin{equation}
\label{eq92}
K_b = - \frac{2\varepsilon _F^0 \rho ^0\left( {\varepsilon _F^0 } 
\right)}{b^0} \cdot \frac{\partial b^\sigma }{\partial m} \quad .
\end{equation}

In further analysis we will use two limiting cases for the on-site Coulomb 
repulsion - weak and strong correlation. 

In the case of {\bf weak correlation} the Coulomb correlation factor $K_U = 
0$, and the critical on-site exchange interaction depends only on the 
inter-site interactions through the inter-site correlation factor $K_b $. 
Using $I_\sigma $ expressed by Eq. (\ref{eq13}) one obtains

\begin{equation}
\label{eq93}
F^{cr} = \frac{\left( {1 - K_b } \right)b^0}{\rho ^0(\varepsilon _F^0 )} - 
zJ - 2D\left[ {\frac{\left( {1 - S} \right)^2}{2}\frac{n^2}{4} + \left( {1 - 
S} \right)\left( {1 - n} \right)} \right] \quad ,
\end{equation}
where the bandwidth change parameter in the paramagnetic state ,$b^0$, is 
given by

\begin{equation}
\label{eq94}
b^0 = 1 - \left( {1 - S} \right)n + \left( {1 - S} \right)^2\left[ 
{\frac{n^2}{4} - 3\left( {\frac{n(2 - n)}{4}} \right)^2} \right] - \frac{2J 
+ J' - V}{t}\frac{n(2 - n)}{4} \quad .
\end{equation}

To illustrate the dependence of the critical on-site Stoner field, $F^{cr}$ 
on electron concentration (at zero temperature) we will use the above 
equations and the initial semi-elliptic DOS given by

\begin{equation}
\label{eq95}
\rho ^0\left( {\varepsilon ^0} \right) = \frac{2}{\pi D}\sqrt {1 - \left( 
{\frac{\varepsilon ^0}{D}} \right)^2} \quad .
\end{equation}

Mathematically this is the simplest possible DOS which can be used for 
itinerant ferromagnetism. The rectangular DOS does not have a unique 
relationship between the Stoner field and the magnetization, and the 
parabolic DOS does not take into account the inter-site interactions, since 
it does not have a finite width and is not centered on the atomic level.

Electron concentration in the paramagnetic state (at T=0K) is calculated 
from the simple condition

\begin{equation}
\label{eq96}
\frac{n}{2} = \int\limits_{ - D}^{\varepsilon _F^0 } {\rho ^0\left( 
{\varepsilon ^0} \right)d\varepsilon ^0} \quad ,
\end{equation}
with expression (\ref{eq95}) inserted for the DOS. This will allow to calculate 
$\varepsilon _F^0 (n)$ and $\rho ^0(\varepsilon _F^0 )$ for a given electron 
occupation $n$. Finally, we calculate $F^{cr}(n)$ from Eq. (\ref{eq93}) with $K_b $ 
estimated from Eq. (\ref{eq92}), with the help of the derivative $\frac{\partial 
b^\sigma }{\partial m}$ calculated from Eq. (\ref{eq22}).

Calculated in this way the dependence of critical on-site Stoner field on 
electron concentration is shown in the Fig. 5 for the case of weak Coulomb 
correlation $U$ and various parameters of inter-site interaction$J$ and $S$, 
at $J' = J$ and $V = 0$. Parameters $S = 1$ and $J = 0$ describe the classic 
Stoner model. 

We can see from this Figure that the decreasing factor $S$ and increasing 
interaction, $J$, decrease the on-site Stoner field required to create 
ferromagnetism, even to zero, at some electron concentrations (solid line). 
Including only the inter-site exchange interaction (dashed line) lowers the 
critical field symmetrically with respect to $n = 1$. The kinetic 
interactions: $t_{ex} $ and $\Delta t$, described by the hopping inhibiting 
factor $S < 1$ (dotted line) lower the on-site Stoner field, but now there 
is no symmetry with respect to the half-filled point $n = 1$. 

\newpage
\begin{figure}[t]
\begin{center}
\epsfxsize11cm
\epsffile{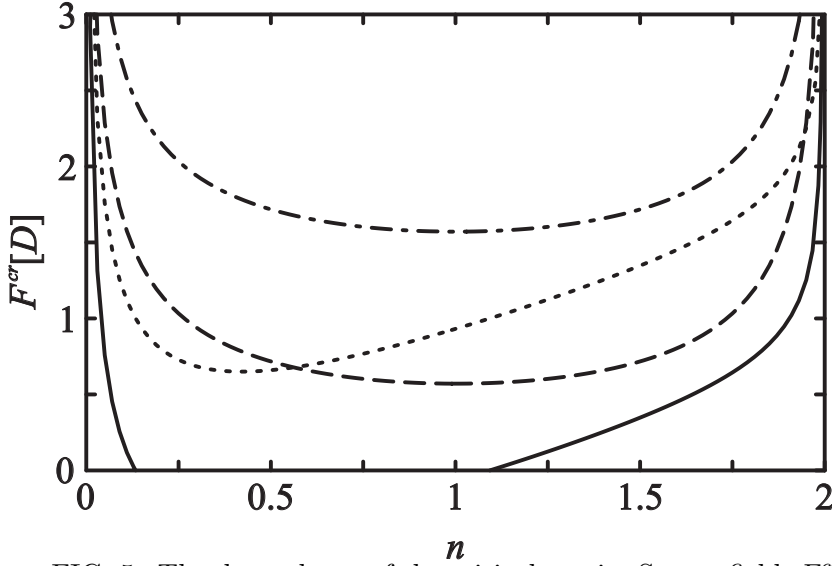}
\caption{The dependence of the critical on-site Stoner field, $F^{cr}$, on the 
electron occupation, for different values of $S$ and $J$; $S = 1$ and $J = 
0$ (the Stoner model) -- dot-dashed line, $S = 0.6$ and $J = 0.5t$ -- solid 
line, $S = 1$ and $J = 0.5t$ -- dashed line, $S = 0.6$ and $J = 0$ -- dotted 
line. This is the case of the weak Coulomb correlation, $U$.}
\label{fig5}
\end{center}
\end{figure}
In {\bf the case of strong correlation} $\left( {U > > D} \right)$ the band 
is split into two sub-bands. For the initial semi-elliptic DOS given by Eq. 
(\ref{eq95}) we obtain from Eq. (\ref{eq30}) in this limit (see \cite{10}) the following 
densities

- for the lower Hubbard sub-band

\begin{equation}
\label{eq97}
\rho _\sigma \left( \varepsilon \right) = \frac{2}{\pi D_{ef} }\sqrt {1 - 
n_{ - \sigma } - \left( {\frac{\varepsilon }{D_{ef} }} \right)^2} \quad ,
\end{equation}
- for the upper Hubbard sub-band

\begin{equation}
\label{eq98}
\rho _\sigma \left( \varepsilon \right) = \frac{2}{\pi D_{ef} }\sqrt {n_{ - 
\sigma } - \left( {\frac{\varepsilon }{D_{ef} }} \right)^2} \quad ,
\end{equation}
with $\varepsilon = \varepsilon ^0b^0$, and $D_{ef} = Db^0$ .

The parameter $I_\sigma $ is given by Eqs (\ref{eq16}) and (\ref{eq17}), which after 
inserting into Eq. (\ref{eq91}) give the following values of the critical on-site 
exchange interaction

-for the lower Hubbard sub-band

\begin{equation}
\label{eq99}
F^{cr} = \frac{\left( {1 - K_U - K_b } \right)b^0}{\rho ^0(\varepsilon _F^0 
)} - zJ - 2D\frac{\left( {1 - n} \right)}{\left( {2 - n} \right)^2}\left[ 
{\frac{\left( {1 - S} \right)^2}{2}n^2 + 4\left( {1 - S} \right)(1 - n)} 
\right] \quad ,
\end{equation}
-for the upper Hubbard sub-band

\begin{equation}
\label{eq100}
F^{cr} = \frac{\left( {1 - K_U - K_b } \right)b^0}{\rho ^0(\varepsilon _F^0 
)} - zJ - 2D\left( {n - 1} \right)\left[ {\frac{\left( {1 - S} \right)^2}{2} 
- 4\left( {1 - S} \right)\frac{\left( {n - 1} \right)}{n^2}} \right] \quad 
,
\end{equation}
where $\rho ^0\left( {\varepsilon _F } \right) = \mathop {\lim }\limits_{m 
\to 0} \rho _\sigma \left( \varepsilon \right) / b^0$, with $\rho _\sigma 
\left( \varepsilon \right)$ given by Eqs (\ref{eq97}), (\ref{eq98}) with $n_{ - \sigma } = 
n / 2$, and $\varepsilon _F (n)$ calculated from Eq. (\ref{eq96}) as $\varepsilon 
_F (n) = \varepsilon _F^0 b^0$. The correlation factor $K_U $ is calculated 
from Eq. (\ref{eq86}), which at T=0K is reduced to the simple condition

\begin{equation}
\label{eq101}
K_U = \int\limits_{ - D_{ef} }^{\varepsilon _F } {\frac{\partial \rho 
_\sigma (\varepsilon )}{\partial m}} d\varepsilon - \int\limits_{ - D_{ef} 
}^{\varepsilon _F } {\frac{\partial \rho _{ - \sigma } (\varepsilon 
)}{\partial m}} d\varepsilon = 2\int\limits_{ - D_{ef} }^{\varepsilon _F } 
{\frac{\partial \rho _\sigma (\varepsilon )}{\partial m}} d\varepsilon \quad 
,
\end{equation}
in which we will use the DOS given by Eqs (\ref{eq97}) and (\ref{eq98}).

We calculate $F^{cr}(n)$ from Eq. (\ref{eq100}), with $K_b $ given by Eq. (\ref{eq92}). The 
bandwidth $b^0$ is calculated from Eq. (\ref{eq22}) with $I_\sigma $ given by the $m 
= 0$ limit of Eqs (\ref{eq20}) and (\ref{eq21}), for the lower and upper sub-band, 
respectively. 

Fig. 6 shows that in the presence of the on-site strong correlation $U$, the 
inter-site interactions: $J,\mbox{ }S$, decrease the critical Stoner field 
dramatically. The increase of $J$ causes the decrease of $F^{cr}$ for 
concentrations nearly half-filled and also at small concentrations and 
concentrations close to completely filled. The decrease of $S$ from 1 to 0 
causes a drop of the critical Stoner field, $F^{cr}$, especially for small 
$n < 0.5$, and for $1 < n < 1.5$, where both Hubbard sub-bands begin to 
fill. As we know the strong Coulomb repulsion causes the split of the band. 
This split causes the change in sign of the inter-site correlation factor 
$K_b $ when the Fermi level moves from the lower to the upper Hubbard 
sub-band. For $S \ne 1$ the ratio $\frac{K_b b^0}{\rho ^0(\varepsilon _F^0 
)} \ne 0$. This is why the curves with $S < 1$will have a discontinuity at 
half filling, when the Fermi energy jumps from the lower to the upper 
sub-band.

\begin{figure}[htbp]
\begin{center}
\epsfxsize11cm
\epsffile{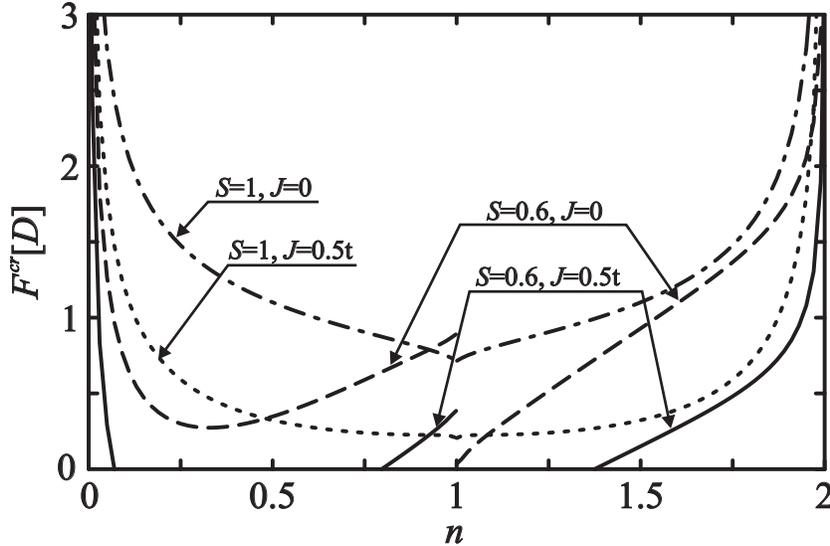}
\caption{Dependence of the on-site critical Stoner field, $F^{cr}$ (in 
the units of half bandwidth), on the electron occupation. The curves show 
the influence of inter-site interactions, $J = J'$, $V = 0$, and of the 
hopping interactions, $\Delta t,\mbox{ }t_{ex} $, represented by the hopping 
`inhibiting' factor $S$ on $F^{cr}$in the presence of strong on-site 
correlation, $U = \infty $.}
\label{fig6}
\end{center}
\end{figure}

To justify the use in this chapter of the rather academic limit of $U > > D$ 
for F ordering we compare on the next graph the three cases; $U = 0$, $U = 
3D$, and $U > > D$. All the curves were calculated on the base of Eqs (\ref{eq95}), 
(\ref{eq30}) and (\ref{eq96}). One can see that the difference between the curves $U = 3D$, 
and $U > > D$ is small. This justifies the use in numerical calculations 
(for simplicity) of the strong correlation limit $U > > D$.

\begin{figure}[htbp]
\begin{center}
\epsfxsize11cm
\epsffile{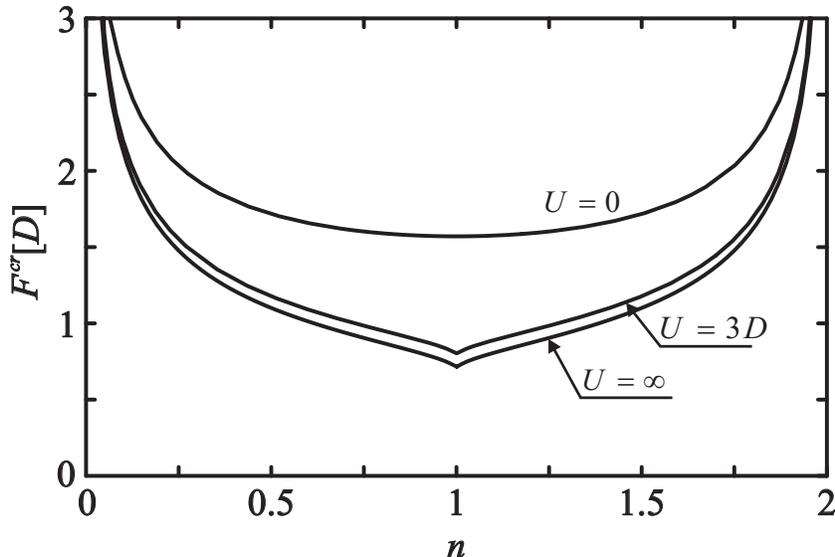}
\caption{Dependence of the on-site critical Stoner field, $F^{cr}$ (in 
the units of half bandwidth), on the electron occupation, which shows the 
influence of on-site correlation, $U$. The inter-site interactions are zero 
($J = 0,S = 1)$}
\label{fig7}
\end{center}
\end{figure}

\vskip0.5cm 

{\bf Enhancement of magnetic susceptibility}
\vskip0.5cm 

The susceptibility given by Eq. (\ref{eq88}) for $T \ne 0\mbox{ K}$ or Eq. (\ref{eq89}) for 
$T = 0\mbox{ K}$ is not divergent for most of the pure elements, as only few 
of them are magnetic. But for many pure elements the denominator of 
susceptibility is significantly decreased, which produces the experimentally 
observed increase of susceptibility. This phenomenon was extensively studied 
in the past. From Eq. (\ref{eq89}) in the H-F approximation for $U$, and in the 
absence of the inter-site interactions, when $K_U \equiv 0$, $K_b \equiv 0$, 
and $b^0 = 1$ (see e.g. \cite{57}) one can write

\begin{equation}
\label{eq102}
\chi = \frac{2\mu _B^2 \rho ^0(\varepsilon _F^0 )}{1 - F_{tot}^F \rho 
^0(\varepsilon _F^0 )} = \chi _P A \quad ,
\end{equation}
where the bare Pauli term $\chi _P = 2\mu _B^2 \rho ^0(\varepsilon _F^0 )$ 
is enhanced by the factor $A = 1 \mathord{\left/ {\vphantom {1 {\left[ {1 - 
F_{tot}^F \rho ^0\left( {\varepsilon _F^0 } \right)} \right]}}} \right. 
\kern-\nulldelimiterspace} {\left[ {1 - F_{tot}^F \rho ^0\left( {\varepsilon 
_F^0 } \right)} \right]}$ the Stoner enhancement factor. This enhancement in 
the case of nonzero correlations is given by 

\begin{equation}
\label{eq103}
A = 1 \mathord{\left/ {\vphantom {1 {\left[ {\left( {1 - K_U - K_b } 
\right)b^0 - F_{tot}^F \rho ^0(\varepsilon _F^0 )} \right]}}} \right. 
\kern-\nulldelimiterspace} {\left[ {\left( {1 - K_U - K_b } \right)b^0 - 
F_{tot}^F \rho ^0(\varepsilon _F^0 )} \right]} \quad .
\end{equation}

In the past the CPA approximation describing the on-site repulsion $U$; $K_U 
\ne 0$, $b^0 = 1$, was used to calculate this enhancement (see \cite{9} for the 
general model of pure elements and e.g. \cite{60,61,62} for susceptibility of 
disordered binary alloys). Experimental data and theoretical results from 
the local density functional method were collected more recently for pure 
transition elements by \cite{56}. In this chapter the factor $A$ (Eq. (\ref{eq103})) is 
additionally increased by the inter-site interactions; $K_b \ne 0$, $b^0 < 
1$. This effect should be also included in the investigation of experimental 
data on the susceptibility enhancement.

\vskip0.5cm 

{\bf 4.2 NUMERICAL RESULTS FOR MAGNETIZATION AND CURIE TEMPERATURE}
\vskip0.5cm 

The Curie temperature, at which magnetization vanishes, is calculated from 
the zero of the susceptibility denominator, Eq. (\ref{eq88}), attained with 
temperature

\begin{equation}
\label{eq104}
1 - K_U - K_b - F_{tot}^F I_T = 0 \quad .
\end{equation}

Quantity $F_{tot}^F $ calculated from this equation at nonzero temperature 
is larger than $F_{tot}^{cr} $, calculated from the same equation (\ref{eq104}) at 
zero temperature, since $I_T $ has its maximum $I_T = {\rho ^0(\varepsilon 
_F^0 )} \mathord{\left/ {\vphantom {{\rho ^0(\varepsilon _F^0 )} {b^0}}} 
\right. \kern-\nulldelimiterspace} {b^0}$ at zero temperature. 

In the case of {\bf weak Coulomb correlation} ($U = 0)$ the Coulomb 
correlation factor, $K_U = 0$, hence the Curie temperature depends only on 
$K_b $ and $I_T $. 

In the case of {\bf strong Coulomb correlation (}$U > > D${\bf ) }according 
to Eq. (\ref{eq104}) we have to include the correlation factor $K_U $ given by Eq. 
(\ref{eq86}). In this case the band is split into two sub-bands. Using the DOS given 
by Eqs (\ref{eq97}) and (\ref{eq98}) we calculate numerically the correlation factor $K_U 
$ from Eq. (\ref{eq101}) where the self-energy was eliminated 

\begin{equation}
\label{eq105}
K_U = 2\int\limits_{ - \infty }^{ + \infty } {\frac{\partial \rho _\sigma 
(\varepsilon )}{\partial m}} f(\varepsilon )d\varepsilon \quad .
\end{equation}

Alternatively, we can calculate directly the magnetization as; $m(T) = 
n_\sigma - n_{ - \sigma } $, with $n_{\pm \sigma } $ from Eq. (\ref{eq36}) for a 
given electron occupation; $n = n_\sigma + n_{ - \sigma } $, and then obtain 
the Curie temperature by searching for the Curie point where$m(T_C ) \to 0$.

The procedure for calculations is the following. We use Eq. (\ref{eq37}) for the 
electron occupation number, which corresponds to a given 3d ferromagnetic 
element in the Table I. We adjust the value of total Stoner field, 
$F_{tot}^F $ , to create the experimental zero temperature magnetization 
$m\left( {T \to 0\mbox{ K}} \right) = n_\sigma \left( {0\mbox{ K}} \right) - 
n_{ - \sigma } \left( {0\mbox{ K}} \right)$. Having obtained $F_{tot}^F $ we 
calculate the on-site Stoner field, $F$, from Eq. (\ref{eq26}) at the given assumed 
inter-site interactions $J$, $S$ and $D = 3$d half band-width according to \cite{63}.

Next, we calculate the Curie temperature with the same constants, $F$, $J$ 
and $S$ using Eq. (\ref{eq38}) for the magnetization and searching for the point 
where $m(T_C ) \to 0$.

The results in a{\bf case of the weak correlation} are collected in the 
Table I.

\begin{table}[tbp]
\caption{Curie Temperatures and Values of On-site Stoner Field for 
Ferromagnetic Elements, Modified Stoner Model with $U = 0$ }
\begin{tabular}
{p{56pt}p{35pt}p{35pt}p{48pt}p{50pt}p{50pt}p{50pt}p{50pt}p{55pt}}
\hline
\raisebox{-1.50ex}[0cm][0cm]{Element}& 
\raisebox{-1.50ex}[0cm][0cm]{$n$}& 
\raisebox{-1.50ex}[0cm][0cm]{$m$}& 
\raisebox{-1.50ex}[0cm][0cm]{$D$[eV]}& 
\multicolumn{4}{p{201pt}}{$T_c \mbox{[K]}$ \par $F(m)\mbox{[eV]}$} & 
\raisebox{-1.50ex}[0cm][0cm]{No. 5  } \\
\cline{5-8} 
 & 
 & 
 & 
 $t[eV]$& 
No.1 \par $S = 0.6$ \par $J = 0.5t$& 
No.2 \par $S = 0.6$ \par $J = 0$& 
No.3 \par $S = 1$ \par $J = 0.5t$& 
No.4 \par $S = 1$ \par $J \equiv 0$& 
$T_c^{exp}$[K] \\
\hline
Fe& 
1.4& 
0.44& 
2.8 \par 0.35& 
2050 \par 1.18& 
3295 \par 3.76& 
3980 \par 2.28& 
4290 \par 4.85& 
1043 \quad  \\
\hline
Co& 
1.65& 
0.344& 
2.65 \par 0.22& 
1690 \par 1.86& 
3300 \par 4.46& 
3880 \par 3.0& 
4710 \par 5.6& 
1388 \quad  \\
\hline
Ni& 
1.87& 
0.122& 
2.35 \par 0.20& 
620 \par 2.33& 
870 \par 4.72& 
1720 \par 4.0& 
1960 \par 6.4& 
627 \quad \\
\hline
\end{tabular}
\label{tab1}
\end{table}

It is very interesting to compare the different results for $T_c $ which are 
shown in this Table. Column no. 4, which is the Stoner model for 
semi-elliptic DOS, shows that all theoretical results of the pure Stoner 
model are much higher than the experimental Curie temperatures (column no. 
5). This means that the Stoner model, which assumes that the on-site atomic 
field creates ferromagnetism, overestimates, to a large extent, the Curie 
temperature. The necessary on-site Stoner field is also unrealistically 
high. Perhaps ferromagnetism is the result of inter-site forces changing the 
bandwidth, since column no. 1 is closest to the experimental results. The 
larger the component of the on-site field (i.e. $S \to 1$, $J \to 0)$, the 
more the $T_c $ results exceed the experimental values. However, one has to 
be careful not to overestimate these inter-site interactions since they 
suppose to be weak. They lower substantially the Curie temperature, but the 
remaining difference between theoretical and experimental values is still 
significant and it has to be attributed to the thermal spin fluctuations, 
since there is an experimental evidence of such fluctuations (see \cite{56,64})

These simple calculations would confirm the earlier attempts of 
understanding the itinerant ferromagnetism as the effect of ordering local 
moments, whose alignment disappear at the Curie temperature. However, the 
moments themselves exist up to the temperatures exceeding the Curie 
temperature by two or three times (see Mizia \cite{65}). We could think about 
local moments as being created by the Stoner on-site field, but their 
ordering would be driven by the inter-site interactions, which is much 
weaker than the on-site field.

In the case of {\bf strong correlation }the procedure of calculations is the 
same as in the case of the weak correlation. First, we adjust the on-site 
exchange interaction $F$(at given inter-site parameters $J,S)$ to the 
magnetic moment at zero temperature. We use the DOS given by Eqs (\ref{eq97}) and 
(\ref{eq98}). The correlation factor is calculated from Eq. (\ref{eq105}). Finally, we 
calculate the Curie temperature from Eq. (\ref{eq104}), or from the condition of 
vanishing magnetization, with the same interactions $F,J$ and $S$. 

Table II shows the results of Curie temperatures obtained for the strong 
on-site correlation from the Stoner model ($S = 1,J = 0)$, and for the cases 
with nonzero kinetic and exchange inter-site interactions. It can be seen 
that the influence of strong on-site correlation ($U = \infty )$ does not 
lower the theoretical Curie temperature when compared to the case of no 
on-site correlation ($U = 0)$ (see Table I), although the necessary on-site 
exchange field, $F$, in the case of strong on-site correlation ($U = \infty 
)$ is lower than without it, $U = 0$ (compare Figs 5 i 6). 

\newpage
\begin{table}[tb]
\caption{Curie Temperatures and Values of On-site Stoner Field for 
Ferromagnetic 3d Elements, Modified Stoner Model with $U = \infty $ }
\begin{tabular}
{p{56pt}p{35pt}p{63pt}p{53pt}p{53pt}p{53pt}p{53pt}p{63pt}}
\hline
\raisebox{-1.50ex}[0cm][0cm]{Element}& 
\raisebox{-1.50ex}[0cm][0cm]{$n$}& 
\raisebox{-1.50ex}[0cm][0cm]{$m$}& 
\multicolumn{4}{p{213pt}}{$T_c 
[\mbox{K}]$ \par $F(m)\mbox{ }[\mbox{eV]}$} & 
\raisebox{-1.50ex}[0cm][0cm]{$T_c^{exp } [\mbox{K}]$ \par } \\
\cline{4-7} 
 & 
 & 
 & 
$S = 0.6$ \par $J = 0.5t$& 
$S = 0.6$ \par $J = 0$& 
$S = 1$ \par $J = 0.5t$& 
$S = 1$ \par $J = 0$  \\
\hline
Fe& 
1.4& 
0.44& 
2420 \par 0.41& 
4000 \par 2.45& 
3810 \par 1.03& 
4960 \par 3.06& 
1043 \quad  \\
\hline
Co& 
1.65& 
0.344& 
1850 \par 1.35& 
3270 \par 3.65& 
3990 \par 1.82& 
5180 \par 4.12& 
1388 \quad  \\
\hline
Ni& 
1.87& 
0.122& 
630 \par 1.94& 
910 \par 4.2& 
1770 \par 2.9& 
2020 \par 5.16& 
627 \quad  \\
\hline
\end{tabular}
\label{tab2}
\end{table}

Now we want to focus on the temperature dependence of the magnetization $m$ 
which is plotted in Fig. 8. The dependence $m(T)$ was calculated for the 
strong correlation and without the correlation for electron occupation 
representing iron. In both cases we assumed; $J = 0$, $S = 1$ and $J = 
0.5t$, $S = 0.6$. 

One can see from this figure that the strong on-site correlation $U$ is not 
helping ferromagnetism at all, as the curves with $U > > D$ lay above the 
corresponding curves with $U = 0$.

\begin{figure}[htbp]
\begin{center}
\epsfxsize11cm
\epsffile{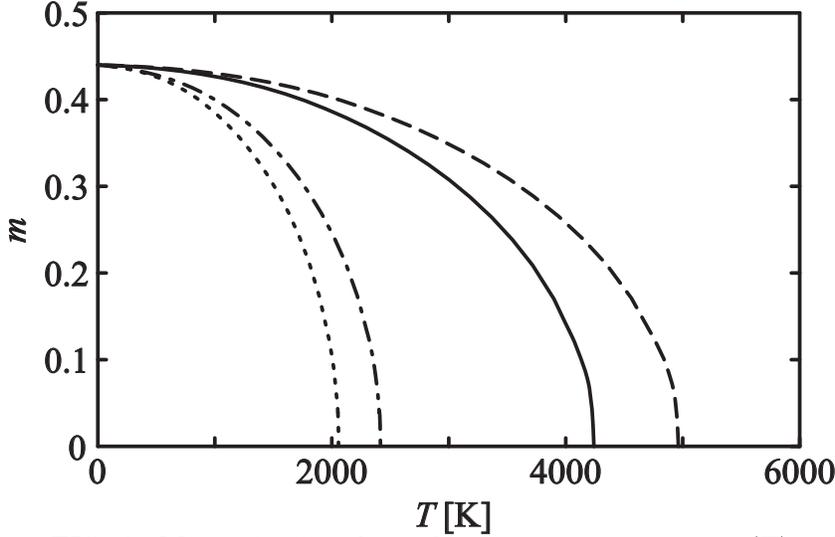}
\caption{Magnetization dependence on temperature, $m(T)$, without and with 
strong Coulomb correlation, $n = 1.4$, $D = 2.8\mbox{ eV}$ for different 
values of parameters $J$ and $S$; $U = 0$, $J = 0$, $S = 1$- solid line, $U 
= \infty $, $J = 0$, $S = 1$- dashed line, $U = 0$, $J = 0.5t$, $S = 0.6$- 
dotted line, $U = \infty $, $J = 0.5t$, $S = 0.6$- dot-dashed line.}
\label{fig8}
\end{center}
\end{figure}

\vskip1cm 

\noindent{\Large{\bf 5 ANTIFERROMAGNETISM}}

\vskip0.5cm 

{\bf 5.1 ONSET OF ANTIFERROMAGNETISM}
\vskip0.5cm 

We can find the critical value of the total field creating AF from Eq. (\ref{eq84}) 
using for $K_x $ the expression (\ref{eq83})

\begin{equation}
\label{eq106}
\begin{array}{c}
 K_x = \int\limits_{ - \infty }^\infty {\frac{\partial \rho _\sigma ^\alpha 
\left( \varepsilon \right)}{\partial x^\sigma }} f\left( \varepsilon 
\right)d\varepsilon = \int\limits_{ - \infty }^\infty {\frac{\partial \rho 
_\sigma ^\alpha \left( \varepsilon \right)}{\partial \Delta }} f\left( 
\varepsilon \right)d\varepsilon \\
= \frac{1}{N}\sum\limits_k {\int\limits_{ - \infty }^\infty {f\left( 
\varepsilon \right)\left( { - \frac{1}{\pi }} \right)} \rm Im \left[ 
{\frac{\partial G_\sigma ^{\alpha \alpha } \left( {\varepsilon ,k} 
\right)}{\partial \Delta }} \right]_{\Delta \to 0} } d\varepsilon  \\
\end{array}.
\end{equation}

Differentiating Eq. (\ref{eq60}) we can show that

\begin{equation}
\label{eq107}
\left. {\frac{\partial G_\sigma ^{\alpha \alpha } \left( {\varepsilon ,k} 
\right)}{\partial \Delta }} \right|_{\Delta \to 0} = \left( 
{\frac{1}{\varepsilon + \varepsilon _k } - \frac{1}{\varepsilon - 
\varepsilon _k }} \right) \cdot \frac{1}{2\varepsilon _k } \quad .
\end{equation}

Using this expression in Eq. (\ref{eq106}) we obtain the following formula

\begin{equation}
\label{eq108}
K_x = - \frac{1}{N}\sum\limits_k {\left[ {\frac{f(\varepsilon _k ) - f( - 
\varepsilon _k )}{2\varepsilon _k }} \right] = \frac{1}{N}} \sum\limits_k 
{\frac{1}{4\varepsilon _k }} \left[ {\tanh \frac{\beta }{2}\left( 
{\varepsilon _k - \mu } \right) - \tanh \frac{\beta }{2}\left( { - 
\varepsilon _k - \mu } \right)} \right] \quad .
\end{equation}

We can calculate now $F_{tot}^{cr} \left( n \right)$ from the relation; 
$F_{tot}^{cr} = 1 / K_x $. The above equation although similar to the 
condition for superconductivity is different in two respects (see \cite{66}). 
First, the summation runs from $k$ minimum in the band to maximum $k$, giving a 
positive $K_x $ and positive $F_{tot}^{cr} \left( n \right)$. Second, in the 
denominator we have the dispersion relation $\varepsilon _k $, which is 
centered on the average energy of the band not around the chemical 
potential, like in the case of superconductivity (see \cite{67}).

From the expression for the electron numbers $n_{\pm \sigma }^\gamma $ 
($\gamma = \alpha ,\beta )$ (Eq. (\ref{eq66})) we obtain in the mean-field 
approximation that the antiferromagnetic moment per atom (in Bohr's 
magnetons) is given by the following expression 

\begin{equation}
\label{eq109}
m = n_\sigma ^\alpha - n_{ - \sigma }^\alpha = \frac{1}{2N}\sum\limits_k 
{\left( {P_k^{ + \sigma } - P_k^{ - \sigma } } \right)\left[ {f(E_k ) - f( - 
E_k )} \right]} \quad .
\end{equation}

The chemical potential $\mu $ is determined from the carrier concentration 
$n $on the basis of the equation, which also comes from Eq. (\ref{eq66})

\begin{equation}
\label{eq110}
n = n_\sigma ^\alpha + n_{ - \sigma }^\alpha = \frac{1}{2N}\sum\limits_k 
{\left( {P_k^{ + \sigma } + P_k^{ - \sigma } } \right)\left[ {f(E_k ) + f( - 
E_k )} \right]} \quad .
\end{equation}

Inserting Eq. (\ref{eq68}) into Eq. (\ref{eq109}) and using the relation (74) we obtain 

\begin{equation}
\label{eq111}
1 = - F_{tot}^{AF} \frac{1}{N}\sum\limits_k {\frac{1}{2E_k }\left[ {f(E_k ) 
- f( - E_k )} \right]} \quad .
\end{equation}

At the transition from AF to normal state; $\Delta \to 0$, and the condition 
(\ref{eq111}) takes on the form

\be
1 = - F_{tot}^{cr} \frac{1}{N}\sum\limits_k {\frac{f(\varepsilon _k ) - f( - 
\varepsilon _k )}{2\varepsilon _k }} = F_{tot}^{cr} \frac{1}{N}\sum\limits_k 
{\frac{1}{4\varepsilon _k }\left[ {\tanh \frac{\beta }{2}\left( {\varepsilon 
_k - \mu } \right) + \tanh \frac{\beta }{2}\left( { - \varepsilon _k - \mu } 
\right)} \right]}
\label{eq118a}
\ee
where $\varepsilon _k = \varepsilon _k^0 b^{AF}$, which is the same as the result (\ref{eq108}) coming from the static 
magnetic susceptibility.

For the numerical analysis we will use the $\Delta \to 0$ limit of Eq. (\ref{eq110}) 
and of Eq. (118) in their integral form,

\begin{equation}
\label{eq112}
n = 2\int\limits_{ - D_{eff} }^{D_{eff} } {\rho \left( \varepsilon 
\right)\,f(\varepsilon )d\varepsilon } \quad ,
\end{equation}

\be
\begin{array}{c}
 1 = - F_{tot}^{cr} \int\limits_{ - D_{eff} }^{D_{eff} } {\rho \left( 
\varepsilon \right)\,\frac{f(\varepsilon ) - f( - \varepsilon 
)}{2\varepsilon }d\varepsilon } \\ 
 = F_{tot}^{cr} \int\limits_{ - D_{eff} }^{D_{eff} } {\rho \left( 
\varepsilon \right)\,\frac{1}{4\varepsilon }\left[ {\tanh \frac{\beta 
}{2}\left( {\varepsilon - \mu } \right) - \tanh \frac{\beta }{2}\left( { - 
\varepsilon - \mu } \right)} \right]d\varepsilon } \\ 
 \end{array} \quad \textrm{, with} \begin{array}{c}
 \varepsilon = \varepsilon ^0b^{AF} \\ 
 D_{eff} = Db^{AF} \\ 
 \end{array} , 
 \label{eq120a}
\ee
and the semi-elliptic DOS given by Eq. (\ref{eq95}) in the case of the weak Coulomb 
correlation $U$. For the strong Coulomb correlation $U$ we will use split 
densities given by Eqs (\ref{eq97}) and (\ref{eq98}). The results will show how the 
critical on-site exchange field depends on the electron concentration. Figs 
9 and 10 show the dependence of this interaction, $F^{cr}$, on the electron 
concentration for different values of the inter-site and kinetic 
interactions, described by the parameters; $J$, $S$, in the case of weak and 
strong Coulomb correlation $U$. Analyzing these curves one can see that the 
inter-site exchange interaction $J$ increases values of on-site exchange 
interaction, $F^{cr}$, required for AF. The decrease in $F^{cr}$ can be 
achieved when $J < 0$. This effect is opposite to the case of 
ferromagnetism, where stabilization of ordering was obtained for $J > 0$. 
Those two different effects of inter-site exchange interaction on magnetism 
can be roughly understood when one compares the expressions for total field 
in the case of ferromagnetism; $F_{tot}^F = F + z[J + t_{ex} \frac{n^2 - 
m^2}{2} + 2\Delta t(1 - n)]$, and antiferromagnetism $F_{tot}^{AF} = F - 
z\left( {J + 2t_{ex} I_{AF} } \right)$. We use the word `roughly' since in 
calculating the critical fields there is included also the factor of the 
band-width change, which does not appear in the effective fields. The result 
of this simplified approach with respect to the inter-site exchange energy, 
$J$, is in agreement with the Heisenberg term for the interaction energy of 
localized spins; $ - J\sum\limits_{i,j} {S_i \cdot S_j } $, which points 
towards ferromagnetism when $J > 0$, and towards antiferromagnetism when $J 
< 0$. Analyzing the influence of inter-site interaction $J$ on long range 
ordering one has to remember that the negative value of this interaction 
stimulates the d-wave superconductivity (see e.g. \cite{68}). As a result it will 
cause the co-existence of SC and AF ordering. This is why the t-J model was 
successful in describing co-existence of the singlet SC and 
antiferromagnetism. Although one has to remember that this model has a 
contradiction in its very origin. This contradiction lies in considering 
electrons as itinerant (the kinetic t-term) and localized (the potential 
J-term) at the same time. Including in our model the kinetic interactions; 
$\Delta t$ and $t_{ex} $, has decreased the minimum on-site exchange 
interaction necessary for AF. The hopping interaction $\Delta t$ decreases 
$F^{cr}$ by decreasing the band-width (see \cite{69}). Positive exchange-hopping 
interaction $t_{ex} $ also decreases the band-width but at the same time it 
decreases the effective field (see above or Eq. (\ref{eq58})). In effect the 
influence of this interaction on AF is very small. 

\newpage
\begin{figure}[tbp]
\begin{center}
\epsfxsize11cm
\epsffile{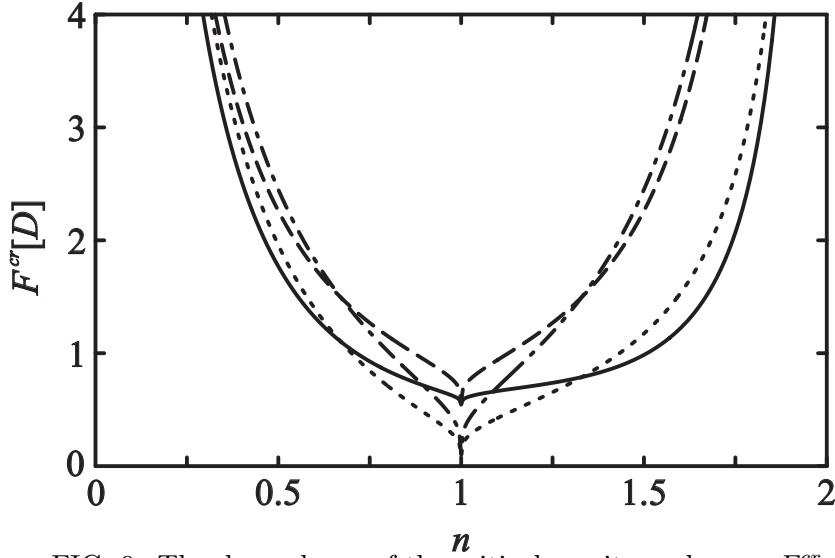}
\caption{The dependence of the critical on-site exchange, $F^{cr}$, on the 
electron occupation, The case of the weak correlation $U$. The curves for 
different values of $S$ and $J$ are; $S = 1$ and $J = 0$-- dot-dashed line, 
$S = 0.6$ and $J = 0.5t$ -- solid line, $S = 1$ and $J = 0.5t$ -- dashed 
line, $S = 0.6$ and $J = 0$ -- dotted line.}
\label{fig9}
\end{center}
\end{figure}

\begin{figure}[htbp]
\begin{center}
\epsfxsize11cm
\epsffile{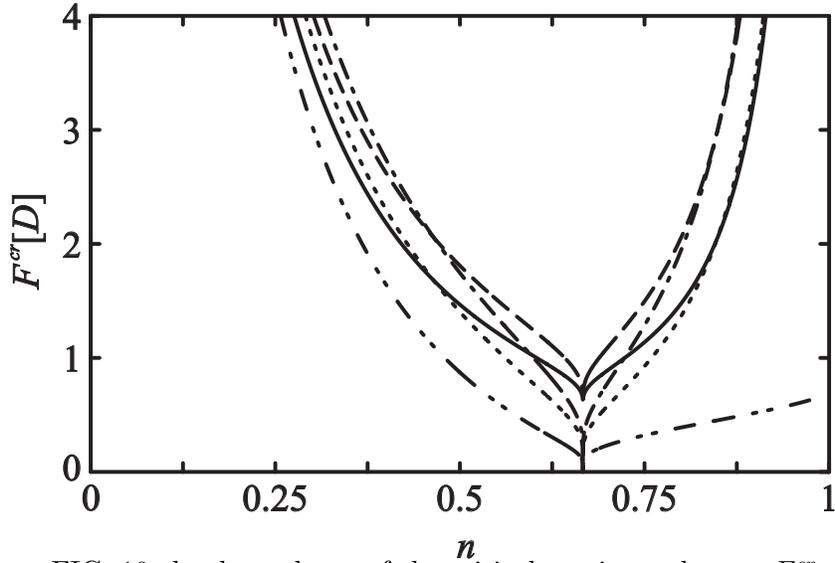}
\caption{he dependence of the critical on-site exchange, $F^{cr}$, on the 
electron occupation. The case of the strong correlation $U$. The curves for 
different values of $S$ and $J$ are; $S = 1$ and $J = 0$-- dot-dashed line, 
$S = 0.6$ and $J = 0.5t$ -- solid line, $S = 1$ and $J = 0.5t$ -- dashed 
line, $S = 0.6$ and $J = 0$ -- dotted line. The double dotted-dashed line is 
for $\Delta t = t$ ($t_1 = 0$ or $S = 0)$, and $t_{ex} \equiv 0$ (see Eq. 
(\ref{eq117}) below).}
\label{fig10}
\end{center}
\end{figure}

Including into calculations the inter-site interactions; $J,J',V$, does not 
shift the minimum of the critical field $F^{cr}(n)$, which for the weak 
correlation is located at $n = 1$, and for the strong correlation at $n = 2 
/ 3$. The inter-site interactions do not change the character of the 
critical curves, they only lower them by a factor of $b^{AF}$ from Eq. (\ref{eq39}) 
or (\ref{eq40}). Remembering that $\varepsilon = \varepsilon ^0b^{AF}$ this can be 
proved analytically 

\begin{equation}
\label{eq113}
\begin{array}{c}
K_x \left( n \right) = - \int\limits_{ - D_{eff} }^{D_{eff} } {\rho \left( 
\varepsilon \right)\,\frac{f(\varepsilon ) - f( - \varepsilon 
)}{2\varepsilon }d\varepsilon } \\
= - \frac{1}{b^{AF}}\int\limits_{- D}^{D} {\rho ^0\left( {\varepsilon ^0} \right)\,\frac{f(\varepsilon ^0b^{AF}) - 
f( - \varepsilon ^0b^{AF})}{2\varepsilon ^0}d\varepsilon ^0} = 
\frac{1}{b^{AF}}K_x^0 \left( n \right) \\
\end{array} \quad .
\end{equation}

Since $F_{tot}^{cr} = 1 / K_x $, and $F_0^{cr} = 1 / K_x^0 $ we obtain that

\begin{equation}
\label{eq114}
F_{tot}^{cr} = F^{cr} - z\left( {J + 2t_{ex} I_{AF} } \right) = 
b^{AF}F_0^{cr} \quad ,
\end{equation}
where $F^{cr}$and $F_0^{cr} $ are the critical on-site exchange fields for 
AF with and without the inter-site interactions, respectively.

Including the kinetic interactions, $S < 1$, leaves the minima of critical 
curves where they are as it only changes the factor $b^{AF}$ in Eq. (\ref{eq114}). 
In this chapter we assumed dependence of the hopping energy on the 
occupation expressed by the equation

\begin{equation}
\label{eq115}
t_{ij}^\sigma = t - \Delta t(\hat {n}_{i - \sigma } + \hat {n}_{j - \sigma } 
) + 2t_{ex} \hat {n}_{i - \sigma } \hat {n}_{j - \sigma } \quad .
\end{equation}

This relation together with the other inter-site interactions; $J,J',V$, 
expressed in the generalized Hartree-Fock approximation brought the Eq. (\ref{eq39}) 
for $b^{AF}$ with $I_{AF} $ given by Eq. (\ref{eq20}) or (\ref{eq21}) in the case of $U > > 
D$. The only exception from the scaling rule for critical on-site exchange 
interaction in the case of $U > > D$ is the situation when $b^{AF}\left( {n 
= 1} \right) = 0$. This would cause the critical curve to drop at $n = 1$, 
what would make antiferromagnetism possible at this concentration. In the 
itinerant band model this is the situation when there is localization at 
half-filling. As a result we obtain at this concentration possible AF, and 
since the band will expand very rapidly with the occupation $n$ departing 
from one there will be also a strong driving force towards SC (see \cite{70}). In 
such circumstances we will have the co-existence or rather competition at $n 
= 1$ between AF and SC. Unfortunately, since$I_{AF} \left( {n = 1} \right) = 
0$, the relation (\ref{eq39}) gives$b^{AF}\left( {n = 1} \right) \ne 0$. In our 
previous paper (see \cite{69}) we have the zero bandwidth at $n = 1$, but we have 
neglected the $t_{ex} $ term in Eq. (\ref{eq3}) or (\ref{eq115}) above. This is equivalent 
to the use of Hirsch's simplified linear approach (see \cite{47,71}) 

\begin{equation}
\label{eq116}
t_{ij}^\sigma = t - \Delta t(\hat {n}_{i - \sigma } + \hat {n}_{j - \sigma } 
) \quad ,
\end{equation}
instead of Eq. (\ref{eq115}){\bf . }From this relation it follows that

\begin{equation}
\label{eq117}
b^{AF} = 1 - \frac{\Delta t}{t}n - \frac{2J + J' - V}{t}I_{AF} \quad .
\end{equation}

At $\Delta t = t$ or $t_1 = 0$ this equation gives$b^{AF}\left( {n = 1} 
\right) = 0$, and we have the possibility of AF ordering, see the double 
dotted-dashed line in Fig. 10. AF can occur at $n = 1$ only when $t_1 \equiv 
0$ what means that there is no hopping in the presence of other electrons 
with opposite spin, and additionally when $t_{ex} \equiv 0$, which leads to 
the strange condition that $t_2 = - t$. Alternatively we can obtain the Eq. 
(\ref{eq116}) (which leads to the AF at $n = 1)$ from the basic Eq. (\ref{eq2}) in a first 
approximation by neglecting all the operator products of the type; $\hat 
{n}_{i - \sigma } \hat {n}_{j - \sigma } $ in Eq. (\ref{eq2}).

\vskip0.5cm  

{\bf 5.2 NUMERICAL RESULTS FOR MAGNETIZATION AND NEEL'S TEMPERATURE}
\vskip0.5cm 

The analysis of Eqs (\ref{eq110}) and (\ref{eq111}) in their integral form will give the 
magnetization dependence on the temperature for different values of the 
on-site and inter-site interactions.

\begin{equation}
\label{eq118}
n = \int\limits_{ - D}^D {\rho \left( \varepsilon \right)\,\left[ {f(E) + f( 
- E)} \right]d\varepsilon } \quad ,
\end{equation}

\begin{equation}
\label{eq119}
1 = - F_{AF} \int\limits_{-D}^D {\rho \left( \varepsilon 
\right)\,\frac{\left[ {f(E) - f( - E)} \right]}{2E}d\varepsilon } \quad ,
\end{equation}
where 

\begin{equation}
\label{eq120}
E = \sqrt {\varepsilon ^2 + \Delta ^2} \quad ,
\quad
\Delta = F_{tot}^{AF} \frac{m}{2} \quad ,
\quad
F_{tot}^{AF} = F - z\left( {J + 2t_{ex} I_{AF} } \right).
\end{equation}

From the above equations we find numerically the sub-lattice magnetization, 
and the Neel's temperature, which is the temperature where this 
magnetization drops to zero. Fig. 11 presents the dependence of Neel's 
temperature, $T_N $, on electron concentration for different values of the 
inter-site and kinetic interactions for {\bf the weak Coulomb correlation}. 
All the curves are for the same on-site exchange field. The curves with 
nonzero inter-site exchange interaction $J$ have a lower Neel's temperature, 
because this interaction decreases the effective exchange interaction; 
$F_{tot}^{AF} $, $F_{tot}^{AF} = F - z\left( {J + 2t_{ex} I_{AF} } \right)$. 
This effect is stronger than the decrease of the bandwidth due to $J$ (see 
Eq. (\ref{eq39})), which increases $T_N $. The assisted hopping interaction $\Delta 
t = t(1 - S)$, and kinetic-exchange interaction $t_{ex} = \frac{t}{2}(1 - 
S)^2$ increase the Neel temperature by decreasing the bandwidth (Eq. (\ref{eq39})).

\begin{figure}[htbp]
\begin{center}
\epsfxsize11cm
\epsffile{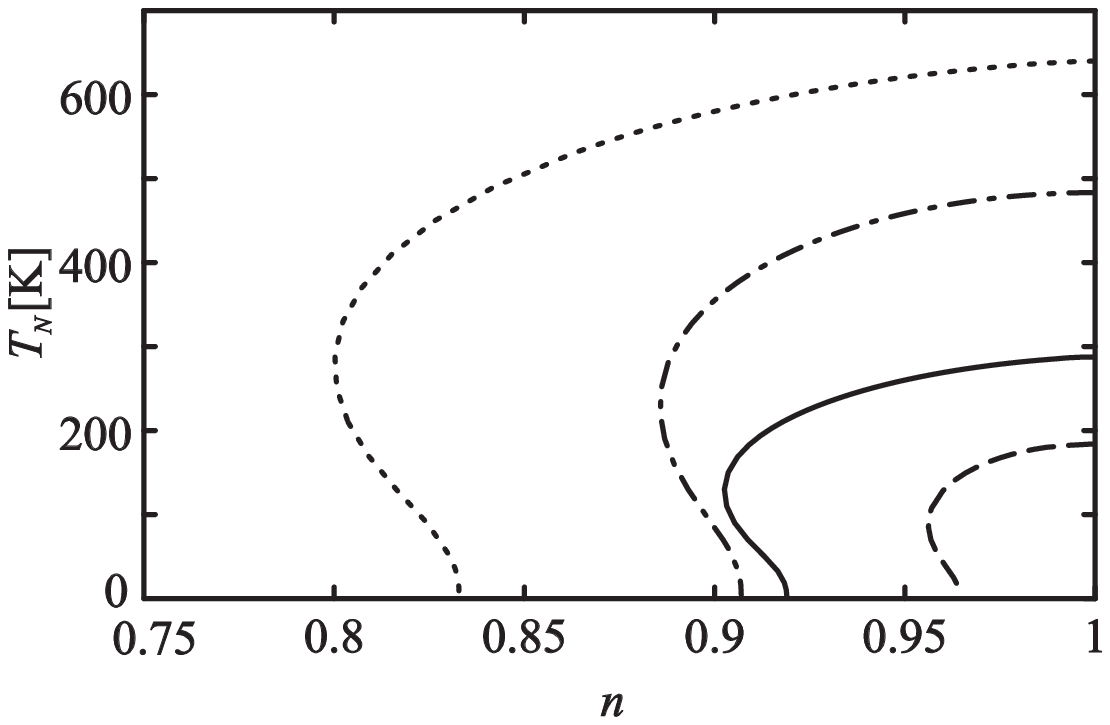}
\caption{Dependence of the Neel's temperature ($T_N )$ on electron 
concentration for $D = 0.5\mbox{ eV}$, $F = 0.34\mbox{ eV}$ and for 
different values of $S$ and $J$in the case of the weak Coulomb correlation; 
$S = 1$ and $J = 0$ -- dot-dashed line, $S = 0.7$ and $J = 0.2t$ -- solid 
line, $S = 1$ and $J = 0.2t$ -- dashed line, $S = 0.7$ and $J = 0$ -- dotted 
line.}
\label{fig11}
\end{center}
\end{figure}

Figure 12 shows the magnetization versus temperature for electron occupation 
$n = 0.95$ and for various values of the inter-site and kinetic 
interactions. Again, all the curves are calculated for the same internal 
exchange field $F = 0.34\mbox{ eV}$. Similarly to the calculations of Neel 
temperature (in Fig. 11) at constant value of the on-site exchange field, 
the highest magnetization is obtained for parameter $S < 1$, which means 
that magnetization is enhanced when electron hopping is inhibited in the 
presence of other electrons. 

\begin{figure}[htbp]
\begin{center}
\epsfxsize11cm
\epsffile{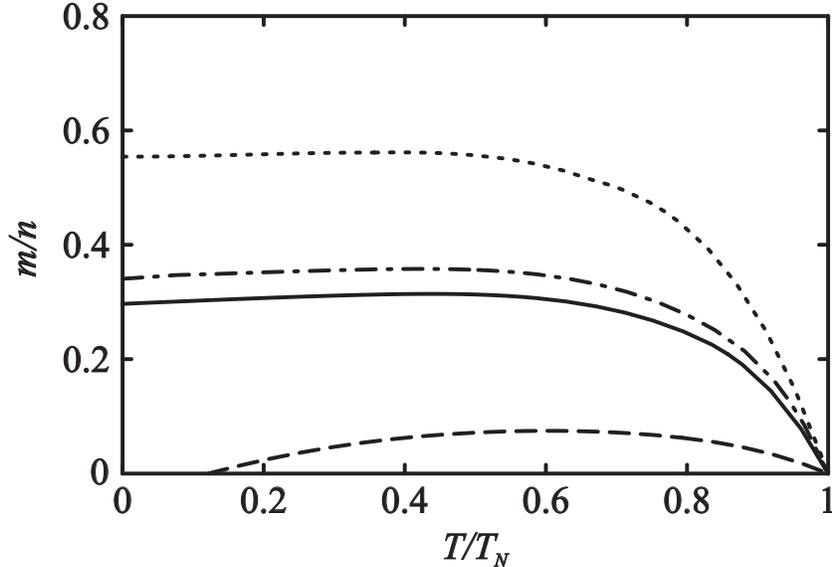}
\caption{ Magnetization versus temperature for band filling $n = 0.96$, $D = 
0.5\mbox{ eV}$, $F = 0.34\mbox{ eV}$ and for different values of $S$ and 
$J$; $S = 1$ and $J = 0$ -- dot-dashed line, $S = 0.7$ and $J = 0.2t$ -- 
solid line, $S = 1$ and $J = 0.2t$ -- dashed line, $S = 0.7$ and $J = 0$ -- 
dotted line. This is the case of the weak Coulomb correlation. }
\label{fig12}
\end{center}
\end{figure}

\begin{table}[tb]
\caption{ {\bf Neel's Temperatures and Values of On-site Stoner Field for Antiferromagnetic Elements}, $U = 0$}
\begin{center}
\begin{tabular}
{p{50pt}p{35pt}p{35pt}p{50pt}p{50pt}p{50pt}p{50pt}p{50pt}p{65pt}}
\hline
\raisebox{-1.50ex}[0cm][0cm]{Element}& 
\raisebox{-1.50ex}[0cm][0cm]{$n$}& 
\raisebox{-1.50ex}[0cm][0cm]{$m$}& 
\raisebox{-1.50ex}[0cm][0cm]{$D$ [eV]}& 
\multicolumn{4}{p{200pt}}{$T_N 
[\mbox{K}]$ \par $F^{AF}(m)\mbox{ [eV]}$} & 
\raisebox{-1.50ex}[0cm][0cm]{No. 5} \\
\cline{5-8} 
 & 
 & 
 & 
 & 
No.1 \par $S = 0.6$ \par $J = 0.5t$& 
No.2 \par $S = 0.6$ \par $J = 0$& 
No.3 \par $S = 1$ \par $J = 0.5t$& 
No. 4 \par $S = 1$ \par $J \equiv 0$&
$T_N^{exp} [K]$  \\
\hline
Cr& 
1.08& 
0.08& 
3.5& 
520 \par 2.37& 
1430 \par 1.47& 
1530 \par 3.17& 
2440 \par 2.27& 
311 \cite{72}\quad  \\
\hline
Mn& 
1.24& 
0.48& 
2.8& 
2110 \par 2.41& 
5840 \par 2.65& 
7090 \par 4.47& 
10820 \par 4.71& 
540 \cite{56}\quad  \\
\hline
\end{tabular}
\label{tab3}
\end{center}
\end{table}

Table III shows the Neel's temperature calculated for the 3d antiferromagnetic 
elements (Cr and Mn) in the case of the weak correlation. First the on-site 
exchange interaction, $F$, at a given assumed inter-site interactions $J$ and 
$S, $ was found from Eq. (\ref{eq43}) after fitting $F_{tot}^{AF} $ to the experimental 
magnetic moment at $T = 0$. Next, using the same $F$ the Neel's temperature 
was calculated. The results show that the inter-site, $J$, and kinetic, $S$, 
interactions decreased the Neel's temperature (column no. 1). The difference 
between experimental and theoretical Neel's temperature is even larger than 
in the case of ferromagnetism (confer Table I). The large difference has to 
be attributed, as in the case of F, to the spin waves or spin fluctuations 
(see \cite{56,64,72}).

\vskip1cm 

\noindent{\Large{\bf 6. CONCLUSION}}

\vskip0.5cm 

We will summarize now the results for ferromagnetism and antiferromagnetism 
obtained in this article. Our analysis of the Hubbard model was focused on 
the influence of the on-site and inter-site Coulomb correlation and the 
kinetic interactions on magnetism. We analyzed magnetic susceptibility, the 
values of critical exchange interaction, the transition temperature and the 
magnetization. 

In analyzing the ferromagnetic ordering we have found three driving forces 
contributing to this ordering. 

(i) spin dependent band shift (Stoner shift) coming from the on-site $F$ and 
inter-site exchange interactions $J$, $V$ and the kinetic interactions 
$\Delta t$, $t_{ex} $ (represented by the parameter $S)$, 

(ii) spin dependent change of the band width and band capacity depending on 
the electron concentrations due to the on-site Coulomb repulsion described 
by the on-site correlation factor $K_U $. Increased capacity of the majority 
spin sub-band drives the magnetism. 

(iii) spin dependent change of the band width and band shape depending on 
the electron concentrations due to the inter-site interactions described by 
the inter-site correlation factor $K_b $ . Narrowing of the majority spin 
sub-band drives the magnetism.

The first effect is the classic effect known since the approach of Weiss 
\cite{73} and Slater \cite{74} to magnetism. This is lowering of the potential 
exchange energy during the transition from the paramagnetic to ferromagnetic 
state. During this transition there is an increase in the kinetic energy. 
The balance of the sum of these two energies decides whether the actual 
transition takes place. This balance leads to the existence of the critical 
values for different interactions, above which the transition takes place. 
In its original form it gives us the Stoner condition for ferromagnetism; 
$F_{cr}^F \rho \left( {\varepsilon _F } \right) > 1$. In the new 
development, described in points two and three above, the increase in the 
kinetic energy is moderated by the on-site and inter-site correlation 
effects. These correlations will also change (decrease) different critical 
interactions coming from the total energy balance. The inter-site and 
on-site correlations are coming into the critical condition throughout the 
correlation factors, which appear in the denominator of the susceptibility 
(Eq. (\ref{eq79})) and as the coefficient of second order term in the free energy 
expansion over magnetization (Eq. (\ref{eq80})), which is proportional to the 
inverse of susceptibility. The inter-site and on-site correlations reduce 
significantly the denominator of the static magnetic susceptibility (Eq. 
(\ref{eq79})). For this reason the magnetic susceptibility is enhanced for the 
materials with strong electron correlation (Eq. (\ref{eq103})). The zeros of the 
denominator give the critical values of the total interaction, which is 
greatly reduced with respect to the classic Stoner model. 

Analysis of the inter-site correlation has shown the great decrease of the 
Curie temperature towards the experimental values (see Table I i II). First, 
the direct calculations within the original Stoner model (see Section 4.2) 
have pointed out that the Curie temperature, after fitting the Stoner shift 
to the experimental magnetic moment at zero temperature, is much too high 
(see Table I). Next, using the modified H-F approximation for the inter-site 
interactions, which is equivalent to introducing the inter-site correlation 
in the first order approximation, we arrived at much lower Curie 
temperatures. Considering the simplicity of the model, and the additional 
presence of spin waves which are out of the scope of this chapter, the 
results were close enough to the experimental data (see Table I). Apparently 
the inter-site interactions are `softer' and decrease faster with the 
temperature than the on-site Stoner field used in the original Stoner model 
(see also \cite{75}). Adding up the on-site strong correlation $U$ to the Stoner 
model do not improve the values of the theoretical Curie temperature (see 
Table II). This fact can be understood better after examining Fig. 6 and Fig. 
7, where the critical field (initializing magnetization) has dropped down at 
half-filling but not at the end of the band, where the 3d elements are 
located.

As already established in the past the electron on-site correlation can help 
in creating {\bf antiferromagnetism} (AF) at half-filled band, where the 
antiferromagnetic 3d elements (Cr, Mn) are located, by dropping the critical 
field for AF to zero (see Fig. 9) in the case of the weak on-site 
correlation. The strong $U$, which gives the correct results for the 
cohesion energy in the middle of the 3d row (see \cite{63,76,77}), shifts the 
zeros of the critical field for AF to the maxima of the two split sub-bands, 
which are located at $n = 2 / 3$, and $n = 4 / 3$ (see Fig. 10). 
Unfortunately, this does not agree with the experimental evidence for AF, 
which is present around $n \approx 1$. The introduction of inter-site 
interactions in the modified H-F approximation does not shift these minima. 
It only lowers the critical curves (see Fig. 10). The inter-site 
interactions ($J,J',V)$ do not reduce the bandwidth at $n = 1$ (which would 
lower the critical curves at this concentration), in the case of the strong 
Coulomb correlation, since $I_{AF} \left( {n = 1} \right) = 0$ (see Eqs (\ref{eq20}) 
and (\ref{eq21})).

There is one exception to this rule. We can obtain AF at half-filling at 
large $U$ if the bandwidth at $n = 1$ goes to zero. This can be achieved in 
the linear approximation, which was described in \S 5.1. The bandwidth was 
reduced to zero by assuming that the hopping integral in the presence of 
another electron is forbidden at concentration $n = 1$; $t_1 = 0$, and 
additionally that $t_{ex} \equiv 0$ (see Eq. (\ref{eq117})).

In conclusion the antiferromagnetism at half-filling in the high correlation 
case can appear when the bandwidth goes to zero at this concentration. It 
can go to zero either as: $1 - n\frac{\Delta t}{t_0 }$, due to the presence 
of the assisted hopping interaction or due to some other interaction, whose 
strength is proportional to $n$.

This situation can describe the high temperature superconducting cuprate 
YBaCuO, where at the half-filling and the strong Coulomb correlation there 
is, initially, the antiferromagnetic order. The superconductivity will 
appear upon doping only away from the half-filling at the electron 
concentration $n \approx 0.95 - 0.8$ (see e.g. \cite{78}).

The temperature dependence of AF was analyzed numerically in Section 5.2. 
Again, as in the case of ferromagnetism the direct calculations within the 
original Stoner model have pointed out that the Neel's temperature, after 
fitting the Stoner shift to the experimental magnetic moment at zero 
temperature, is much too high (see Table III). Next, using the modified H-F 
approximation for the inter-site interaction, we arrived at much lower Curie 
temperatures. They are not as close to the experimental data as in the case 
of ferromagnetism. The reason for this could be the existence of non linear 
modes of excitations in Chromium and Magnesium (see e.g. \cite{72}). 
Nevertheless, as already mentioned above, the model is very simple, the 
details of the realistic DOS as well as the magnetization decrease through 
the spin waves excitation or moment fluctuations could be included, which 
would bring the theoretical results to complete agreement with the 
experimental data.

\end{document}